%% file: main.tex
\documentclass[twocolumn]{aastex631}

\newcommand{\TheStar}{J0931$+$0038\xspace}
\newcommand{\GaiaID}{\gaia DR3 3841101888330639872}

\usepackage{color}
\usepackage{soul}
\usepackage{amsmath}
\usepackage{xspace}
\usepackage{graphicx}
\usepackage{enumitem}
\usepackage{amssymb}
\usepackage{xifthen}
\usepackage{hyperref}
\usepackage[normalem]{ulem}
\usepackage{multirow}

\hypersetup{    
  colorlinks      = {true},
  linkcolor       = {blue},
  citecolor       = {blue},
  urlcolor        = {blue},
}

\newcommand{\code}[1]{\texttt{#1}\xspace}

\newcommand{\gaia}{\textit{Gaia}\xspace}

\newcommand{\unit}[1]{\ensuremath{\mathrm{\,#1}}\xspace}
\newcommand{\feh}{\unit{[Fe/H]}}
\newcommand{\teff}{\ensuremath{T_\mathrm{eff}}\xspace}
\newcommand{\logg}{\ensuremath{\log\,g}\xspace}
\newcommand{\alphafe}{\unit{[\alpha/Fe]}}

\newcommand{\kms}{\unit{km\,s^{-1}}}

\newcommand{\msun}{\unit{M_\odot}}

\newcommand{\minesweeper}{\texttt{MINESweeper}\xspace}

\usepackage{pifont}
\newcommand{\cmark}{\ding{51}}%
\newcommand{\xmark}{\ding{55}}%

\usepackage{xcolor}
\definecolor{ForestGreen}{RGB}{0,155,85}
\definecolor{OliveGreen}{RGB}{60,128,49}
\definecolor{Dandelion}{RGB}{250,162,26}
\definecolor{YellowGreen}{RGB}{152,204,112}
\newcommand{\cxmark}{\textsf{\textcolor{red}{\xmark}}\xspace}
\newcommand{\ccmark}{\textsf{\textcolor{OliveGreen}{\cmark}}\xspace}
\newcommand{\cxmarkq}{\textsf{\textcolor{Dandelion}{\xmark?}}\xspace}
\newcommand{\ccmarkq}{\textsf{\textcolor{YellowGreen}{\cmark?}}\xspace}

\shorttitle{Spectacular Star}
\shortauthors{Ji et al.}

\begin{document}

\title{Spectacular nucleosynthesis from early massive stars}

\input{authors}

\correspondingauthor{Alexander~P.~Ji}
\email{alexji@uchicago.edu}



\begin{abstract}
Stars formed with initial mass over 50 $M_\odot$ are very rare today, but they are thought to be more common in the early universe. 
The fates of those early, metal-poor, massive stars are highly uncertain. Most are expected to directly collapse to black holes, while some may explode as a result of rotationally powered engines or the pair-creation instability. 
We present the chemical abundances of J0931+0038, a nearby low-mass star identified in early followup of SDSS-V Milky Way Mapper, which preserves the signature of unusual nucleosynthesis from a massive star in the early universe. 
J0931+0038 has relatively high metallicity ([Fe/H] $= -1.76 \pm 0.13$) but an extreme odd--even abundance pattern, with some of the lowest known abundance ratios of [N/Fe], [Na/Fe], [K/Fe], [Sc/Fe], and [Ba/Fe]. 
The implication is that a majority of its metals originated in a single extremely metal-poor nucleosynthetic source.
An extensive search through nucleosynthesis predictions finds a clear preference for progenitors with initial mass $> 50 M_\odot$, making J0931+0038 one of the first observational constraints on nucleosynthesis in this mass range. However the full abundance pattern is not matched by any models in the literature.
J0931+0038 thus presents a challenge for the next generation of nucleosynthesis models and motivates study of high-mass progenitor stars impacted by convection, rotation, jets, and/or binary companions.
Though rare, more examples of unusual early nucleosynthesis in metal-poor stars should be found in upcoming large spectroscopic surveys.
\end{abstract}



\section{Introduction} \label{sec:intro}

The chemical abundances of metal-poor stars provide an archaeological snapshot of the first massive stars \citep[e.g.,][]{Frebel15}. When those stars died, they ejected elements that polluted the interstellar and intergalactic medium. Stars forming out of this minimally polluted gas would be metal-poor, and the low-mass metal-poor stars could survive until today, where they can be found in our Milky Way. The atmospheres of these low-mass stars thus provide a window to nucleosynthesis in the first massive stars.
Since even JWST is unable to directly observe the first massive stars \citep[e.g.,][]{Schauer2020}, these chemical abundances are one of the few ways to understand how the first stars formed and died.
Theoretically, one of the most robust predictions is that the first metal-free stars should have a top-heavy initial mass function with characteristic mass $\gtrsim 10\msun$ \citep[e.g.,][]{Bromm13,Klessen2023}.  This prediction, however, is still not confirmed observationally, nor is there a clear understanding of when or how the initial mass function transitions to its present-day shape \citep[e.g.,][]{Offner2014,Sharda2022}.

Decades of searches have led to the discovery and chemical characterization of hundreds of extremely metal-poor stars with $\mbox{[Fe/H]} \lesssim -3$ \citep[e.g.,][]{Beers1992,Cayrel04,Frebel06a,Schlaufman2014,Aguado2016,Starkenburg2017,Li2018,DaCosta2019}.
Their chemical compositions reveal a variety of processes occurring in the early universe.
The majority of these metal-poor stars broadly look like they have been enriched by core-collapse supernovae, possibly following a standard Salpeter initial mass function \citep[e.g.,][]{Cayrel04,Heger10}.
A prominent signature is the carbon-enhanced metal-poor stars, which make up the majority of stars at $\mbox{[Fe/H]} \lesssim -4$ \citep[e.g.,][]{Norris13,Placco14} and may suggest the first stars preferentially explode as faint supernovae \citep[e.g.,][]{Umeda02} or have extremely rapid rotation and winds \citep[e.g.,][]{Meynet06,Chiappini2013}.
A few iron-poor stars have been found without carbon enhancement, making them stars with the lowest overall metallicities \citep[e.g.,][]{Caffau13,Starkenburg2017}.
There also have been many signatures of high energy hypernovae, accompanied by a variety of neutron-capture nucleosynthesis signatures \citep[e.g.,][]{Ezzeddine2019,Yong2021,Skuladottir2021}.
Recently, the first signature of pair instability supernovae has finally been found (\citealt{Xing2023}, though a core-collapse supernova interpretation has also been suggested by \citealt{Jeena2023}).

Interestingly, when comparing the abundances of metal-poor stars to nucleosynthesis models, almost all the supernova progenitors have initial masses less than 50 $\msun$ \citep[e.g.,][]{Placco2015,Fraser2017,Ishigaki2018}.
This could be because more massive stars typically collapse directly to black holes, either directly or after pair-instability pulsation-driven mass loss, and thus do not release any metals into the universe \citep[e.g.,][]{Heger2003, Yoon2012}.

The community's attention has primarily been focused on extremely metal-poor stars with $\mbox{[Fe/H]} \lesssim -3$.
Such metal-poor stars are likely enriched by only a few, or even just one supernova \citep{Audouze95,Ryan96}, so it is reasonable to compare their chemical abundances to nucleosynthesis models of individual supernova explosions.
However in principle, it is possible to find stars dominated by nucleosynthesis in a small number of supernovae at higher metallicities.
For example, it is well-known that pair instability supernovae (PISNe) produce so much calcium and iron that they immediately enrich stars to $\mbox{[Fe/H]} \sim -2$, which would make them difficult to discover in surveys looking for the most Fe-poor stars \citep{Karlsson2008, Salvadori2019}.
Identifying such relatively metal-rich stars with unique elemental compositions is difficult, because the vast majority of stars at $\mbox{[Fe/H]} > -3$ have experienced ordinary chemical evolution, so it is hard to distinguish interesting stars from a vast background of ordinary stars. Stars at higher metallicities could be hiding signatures of a different population of supernovae that produce large amounts of iron.

Here, we present the discovery and chemical composition of the spectacular star 2MASS~J09311004+0038042 (\GaiaID, abbreviated as \TheStar), which was identified in early SDSS-V data.
\TheStar has a relatively high metallicity $\mbox{[Fe/H]} = -1.76$, but its extreme low abundances of other elements like Na, K, Sc, and Ba show it is dominated by nucleosynthesis from a single source.
The star's composition is unlike any star that has been seen before, and its high metallicity and abundance pattern imply a progenitor star with initial mass over 50 \msun, one of the first and most complete observational constraints on nucleosynthesis in this mass range.
However, we have been unable to find satisfactory nucleosynthesis models to explain the full abundance pattern.
Section~\ref{sec:obs} describes our target selection, observations, and chemical abundance analysis.
Section~\ref{sec:results} compares the results of the abundance analysis to existing stellar abundances.
We discuss the origin of this star in Section~\ref{sec:discussion} and
conclude in Section~\ref{sec:conclusion}.
An extended Appendix provides details on the abundance analysis (Appendix~\ref{app:sp})
and nucleosynthesis fits (Appendix~\ref{app:nuc}).

\section{Observations and Analysis} \label{sec:obs}

\begin{figure*}
    \centering
    \includegraphics[width=0.9\linewidth]{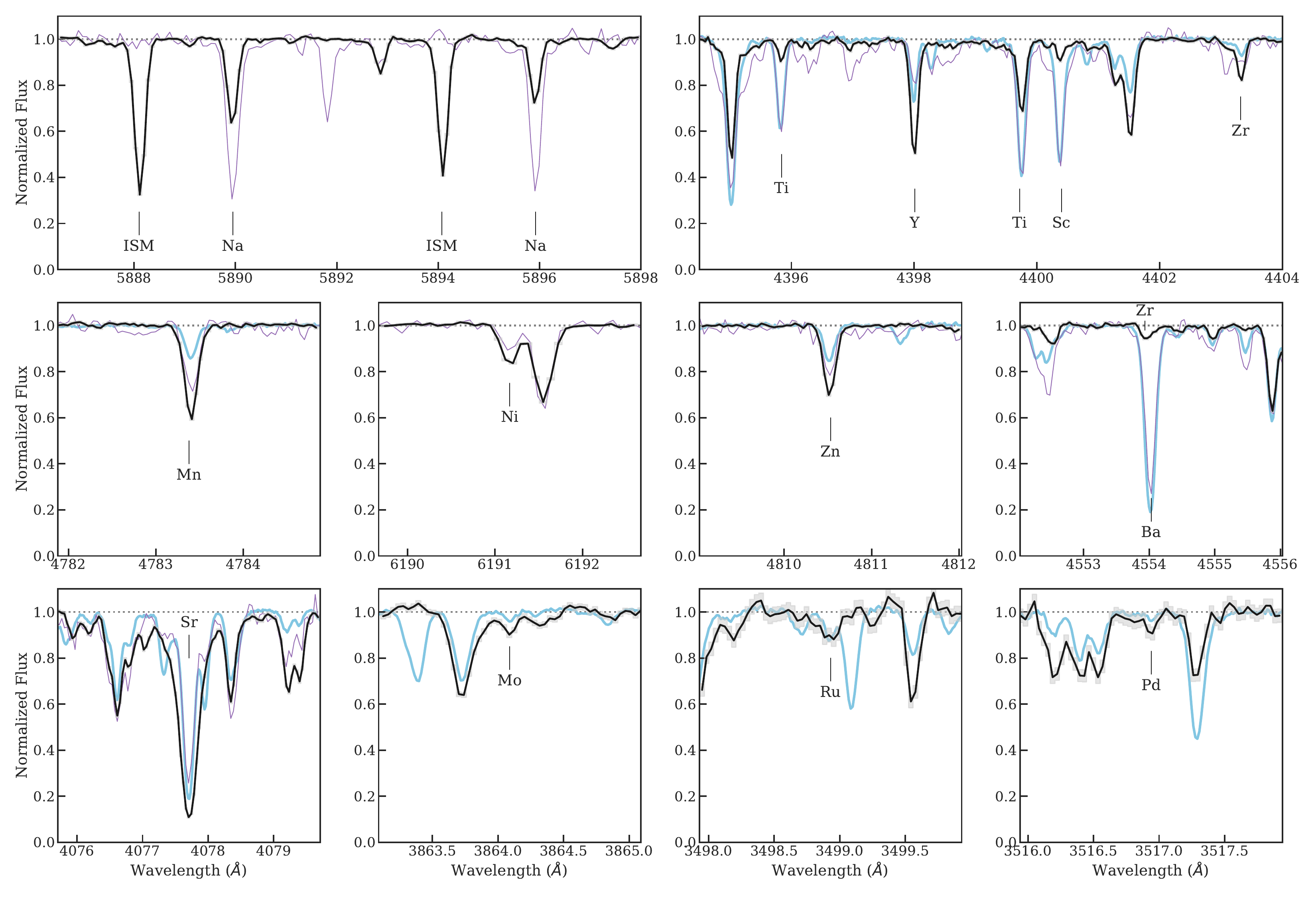}
    \caption{Spectrum of \TheStar (black line and grey shaded uncertainty band).
    For comparison, two stars with similar stellar parameters are plotted: a MIKE spectrum of a normal star from our SDSS sample in purple, and a Keck/HIRES spectrum of the r-process enhanced star BD$+17^\circ3248$ in blue.
    Visible by eye are the low abundances of Na, Ti, Sc, and Ba; the enhancement in Sr, Y, Mn, Ni, and Zn; and the detected Mo, Ru, and Pd lines. 
    BD$+17^\circ3248$ is highly r-process enhanced, but \TheStar has stronger lines of Sr, Mo, Ru, and Pd while the Ba line is barely detected (it is blended with a weak Zr line).
    }
    \label{fig:spectra}
\end{figure*}

\subsection{Target Selection and Observations}

\TheStar was observed by SDSS-V \citep[][De Lee et al., in prep]{Kollmeier2017,Almeida2023} in the metal poor halo program, which uses spare fibers to target stars with photometric metallicities $\mbox{[Fe/H]} < -2$.
This particular star was identified as a metal-poor candidate with SkyMapper DR2 photometry \citep{Onken2019} and observed with the low-resolution optical BOSS spectrograph.
The BOSS spectra were analyzed using \minesweeper{} \citep{Cargile2020}, which performs a spectrophotometric fit including the Gaia DR3 parallax, broadband photometry, and the MIST isochrones \citep{Choi2016}.
The \minesweeper{} parameters were $\teff=5220$K, $\logg=2.57$, $\feh=-1.9$, and $\alphafe = 0.03$, showing it to be a metal-poor and alpha-poor red giant. The star has an eccentric halo orbit and is likely unassociated with any known structures (see Appendix~\ref{app:sp}).

We observed \TheStar with Magellan/MIKE \citep{Bernstein03} for three hours on 2023 April 13, obtaining a high signal-to-noise $R \sim 30,000$ spectrum (100/pixel or 70/resolution element at 4000{\AA}).
The data were reduced with CarPy \citep{Kelson03}.
Portions of this spectrum are shown in Fig~\ref{fig:spectra}, compared to two stars of similar stellar parameters and metallicities: a Keck/HIRES spectrum of the r-process enhanced star BD$+17^\circ3248$ \citep{Johnson02b,Cowan02}; and a Magellan/MIKE spectrum of Gaia DR3 3963318275114883584, a star with ordinary composition.
Just visually, the spectrum of \TheStar displays extraordinarily weak Na, Ti, Sc, and Ba lines; unusually strong lines of Sr, Y, Mn, Ni, and Zn; and clear detections of Mo, Ru, and Pd.

\subsection{Abundance Analysis}

We performed a standard analysis using 1D ATLAS model atmospheres \citep{Castelli04} and the MOOG radiative transfer code including scattering \citep{Sneden73, Sobeck11} and assuming local thermodynamic equilibrium (LTE)\footnote{\url{https://github.com/alexji/moog17scat}}.
The line list was selected from a combination of lines from \citet{Roederer2018} and \citet{Ji2020}, with atomic data adopted from \code{linemake}\footnote{\url{https://github.com/vmplacco/linemake}} \citep{Placco2021}.
Stellar parameters were derived by fixing spectrophotometric temperatures and then determining other stellar parameters spectroscopically, resulting in $\teff = 5200 \pm 100$K, $\logg = 2.75 \pm 0.20$,  $\nu_t = 1.65 \pm 0.3 \kms$, $\mbox{[M/H]}=-1.9 \pm 0.1$, and $\alphafe=0.0$.
Chemical abundances were determined using \code{smhr} \citep{Casey14}\footnote{\url{https://github.com/andycasey/smhr}} with a mix of equivalent widths and syntheses.
Upper limits were calculated using synthetic spectra.
The adopted abundance uncertainty includes line-to-line scatter, signal-to-noise, and stellar parameter uncertainties. [Fe/H] uncertainties are on the total metallicity, while [X/Fe] uncertainties are relative to [Fe/H].
Non-LTE (NLTE) corrections were mostly calculated using \code{TSFitPy}\footnote{\url{https://github.com/TSFitPy-developers/TSFitPy}} \citep{Gerber2023}.
The actual [Fe/H] abundance after NLTE corrections is $\mbox{[Fe/H]} = -1.76 \pm 0.13$.
We also estimated evolutionary corrections for C and N based on metal-poor giants in APOGEE DR17 \citep{Abdurrouf2022}.
Full details of the analysis are given in Appendix~\ref{app:sp}.

\begin{figure*}
    \centering
    \includegraphics[width=\linewidth]{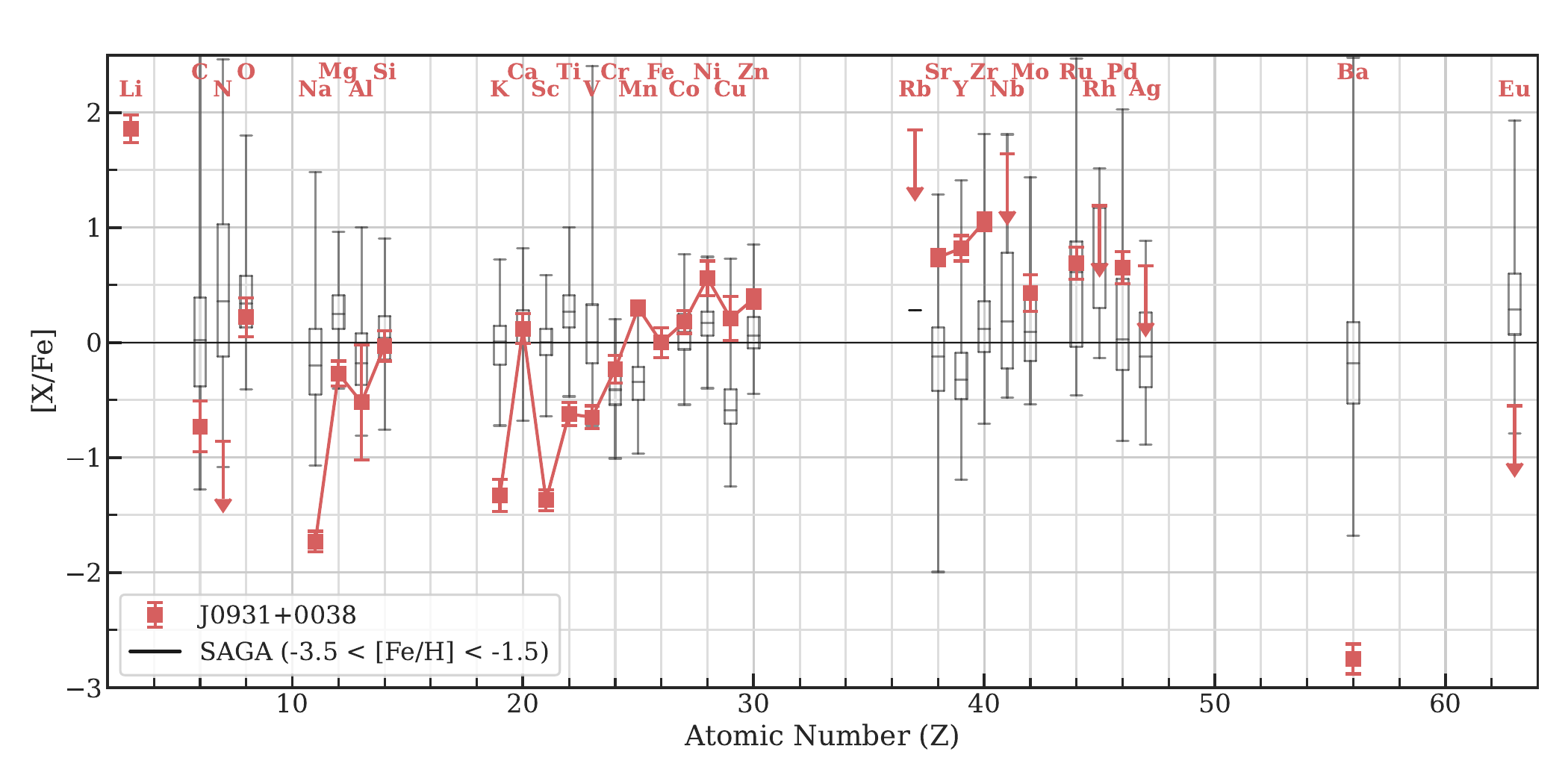}
    \caption{[X/Fe] (in NLTE when possible) vs atomic number.
    The grey boxplots for each element $X$ are [X/Fe] from 4866 stars in the SAGA literature compilation \citep{Suda2008} with $-3.5 < \mbox{[Fe/H]} < -1.5$, to maximize the intrinsic [X/Fe] range. The SAGA abundances have been moved to the \citet{EMagg2022} abundance scale, and they have been shifted by the NLTE correction of \TheStar in Table~\ref{tab:abunds}.
    The box-and-whisker plots indicate the median and 25th-75th percentile with the box and the 1st-99th percentile with the whiskers.
    The abundance of Na, K, Sc, and Ba are among the lowest abundances of these elements ever measured.
    The uncertainty for Fe is the overall metallicity uncertainty, while for other elements it is the precision relative to Fe. Note the plotted Li value is [Li/Fe], and A(Li)=1.15.
    }
    \label{fig:xfe_saga}
\end{figure*}

\section{Abundance Results} \label{sec:results}

Table~\ref{tab:abunds} presents the chemical abundances of \TheStar, in NLTE where available.
Figure~\ref{fig:xfe_saga} shows [X/Fe] compared to the SAGA database after removing upper limits \citep{Suda2008}.
We adopt the Solar abundance scale from \citet{EMagg2022}, using \citet{Asplund09} to fill in missing elements. The SAGA database abundances are shifted to this abundance scale.
The SAGA database predominantly consists of LTE abundances, so for comparison we also shift the SAGA abundances by the NLTE corrections for \TheStar in Table~\ref{tab:abunds}.

There are four remarkable features in the abundance pattern of \TheStar.
First, the light elements from C to Sc display an extremely strong odd-even effect, comparable only to the recently discovered ``pair instability'' star J1010$+$2358 \citep{Xing2023}.
Second, the abundances of the light iron-peak elements Sc, Ti, and V are extremely low, similar to some metal-poor stars in the bulge, halo, and dwarf galaxies \citep{Casey15, Ji2020a}.
Third, the heavier iron-peak elements Mn, Ni, and Zn are quite enhanced, which matches some extremely metal-poor stars associated with hypernovae \citep{Ezzeddine2019,Yong2021,Skuladottir2021}.
Fourth, the neutron-capture elements around the first peak (magic neutron number $N=50$) from Sr to Pd are highly enhanced similar to stars like HD122563 and HD88609 \citep{Honda07}, but the [Ba/Fe] is one of the lowest values ever measured, comparable to the most extreme stars in ultra-faint dwarf galaxies \citep{Ji19a}.
While each of these four features has been seen before in individual stars, \TheStar displays one of the most extreme versions of each feature and combines all of them in one star.
Additionally, all previously known stars with such extreme abundance features have been found in the very metal-poor regime at $\mbox{[Fe/H]} \lesssim -2.5$, but \TheStar has a metallicity over 5${\times}$ higher $\mbox{[Fe/H]} = -1.76$.

\input{abundtable}

\section{Discussion}\label{sec:discussion}

\subsection{A Single Enrichment Source}\label{sec:dilution}

\begin{figure}
    \centering
    \includegraphics[width=0.85\linewidth]{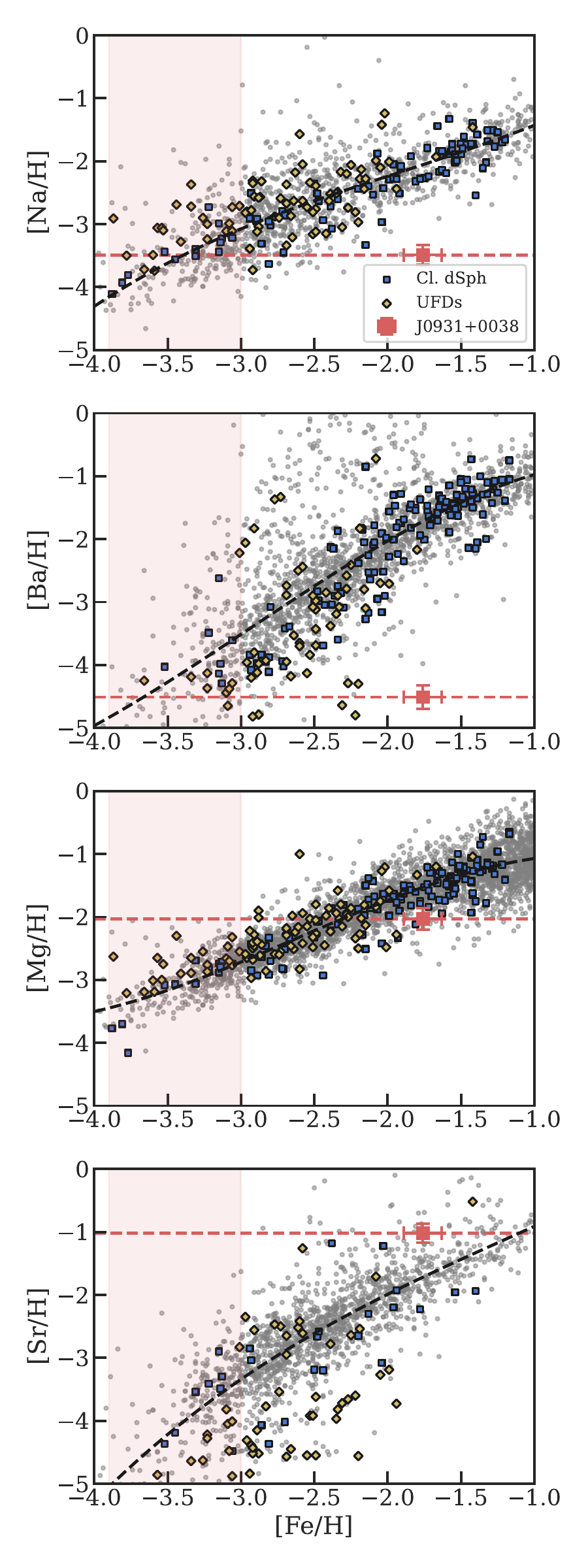}
    \caption{
    Composition of \TheStar (red square) for Na, Ba, Mg, and Sr compared to the SAGA database (grey points), higher mass classical dwarf galaxies (blue squares), and lower mass ultra-faint dwarf galaxies (yellow diamonds).
    The dashed black line is an outlier-clipped 3rd order polynomial fit to the SAGA data.
    The horizontal red line is the [X/H] of \TheStar.
    From the top two panels, the red shaded region indicates $\mbox{[Fe/H]} < -3$, the ISM metallicity range that alone would contribute all of the Na, Ba, and other under-abundant elements like N, K, and Sc observed in \TheStar.
    The bottom two panels, Mg and Sr, show that explaining the low N, Na, K, Sc, and Ba by simply adding Fe (e.g. with a Type Ia supernova) would require extremely high abundances of other elements.
    }
    \label{fig:fehxh}
\end{figure}

\TheStar has a relatively high metallicity and would normally be considered heavily contaminated by chemical evolution, but the very low abundances of N, Na, K, Sc, and Ba imply that it has negligible contamination from general chemical evolution in the interstellar medium (ISM). 
Figure~\ref{fig:fehxh} illustrates this, by plotting [X/H] vs [Fe/H] of \TheStar and the SAGA database, including dwarf galaxy stars from JINAbase (\citealt{jinabase}; other elements shown in Appendix~\ref{app:sp}).
It is clear that \TheStar is not part of any overall chemical evolution trend, either in Milky Way halo stars or in all known dwarf galaxies.

A newly formed star's metallicity is the sum of the metallicities of the background ISM and the diluted ejecta from any recent nucleosynthetic sources.
Thus, one way to interpret Figure~\ref{fig:fehxh} is that 100\% of the Na and Ba in \TheStar comes from swept up ISM material. This would imply that it formed from ISM composition of $\mbox{[Fe/H]} \sim -3.5$ (red dashed lines in Figure~\ref{fig:fehxh}).
A recent nucleosynthetic event would have to raise the ISM metallicity by a factor of $\gtrsim$50 from $\mbox{[Fe/H]} \sim -3.5$ to $\mbox{[Fe/H]} = -1.76 \pm 0.13$, without adding \emph{any} N, Na, K, Sc, or Ba.
Thus, the extreme low abundance of these elements implies that the majority of metals in \TheStar have to be made in ``one shot'', i.e. from a single nucleosynthetic event, rather than the continuous sum of multiple sources as expected in ordinary chemical evolution.
This is a conservative interpretation, because if the nucleosynthetic event produced any N, Na, K, Sc, or Ba, the ISM would have been even lower metallicity, possibly even primordial composition.

The presence of multiple extreme abundance ratios in \TheStar also favors a single source of elements, rather than combining multiple stellar sources. Each extreme ratio is erased by mixing with ordinary ISM, so invoking multiple element sources requires spatial and temporal coincidence, as the homogenization time in dwarf galaxies is only ${\sim}100-300$ Myr \citep[see references in][]{Ji2023}.
For example, it is tempting to invoke a Type~Ia supernova in combination with a massive star supernova to explain the high metallicity, low odd-even, and unusual Fe peak abundances (in an analogy to the ``iron-rich metal-poor stars,'' \citealt{Reggiani2023}).
The C-Ca and neutron-capture elements in \TheStar can not come from the Type~Ia, so must instead originate from ISM material mixed with the Type~Ia ejecta. However, this ISM would need to have one of the highest [Mg, Si, Ca/Fe] abundances ever observed, as well as simultaneously the highest [Sr, Y, Zr/Fe] abundances known.
This is illustrated in Fig~\ref{fig:fehxh}: of the elements shown, only Fe is produced in Type~Ia SNe, so the horizontal red lines indicate the track of Type~Ia SN enrichment. If \TheStar originated from the shaded red [Fe/H] region and was enriched to high [Fe/H] by a Type~Ia SN, it must have had the highest [Mg/Fe] and [Sr/Fe] ever observed.
Thus, a Type~Ia supernova can only be invoked if it occurs simultaneously in the same region of a galaxy as an extreme core-collapse supernovae that produced the high abundance of Mg, Sr, and other elements, which would be an implausible coincidence.
A similar argument precludes most other combinations of multiple sources, though it may be plausible to combine two core-collapse supernovae that originate from the same binary system.

\subsection{Maximum metallicity of supernova models}\label{sec:highfe}
An extraordinary nucleosynthetic event is needed to produce the high metallicity of \TheStar in one shot.
We can constrain this by modeling the maximum metallicity achievable from stars forming directly out of a supernova explosion mixed with pristine gas.
A supernova with a given explosion kinetic energy will sweep up a minimum mass of gas before the material can turn into stars \citep[e.g.,
][]{Cioffi88,Ryan96,Macias2018,Ji2020a,Magg2020,Kolborg2022}.
This imposes an \emph{upper limit} on [Fe/H]: it is possible to dilute the supernova metal yield into more gas, but not less.

To compare this to \TheStar, we take the explosion energy and iron yield for a wide range of supernova nucleosynthesis models covering different progenitor masses, fallback, energies, and metallicities up to $\mbox{[Z/H]}<-1.5$ \citep{Heger2002ApJ,Heger10,Nomoto2013,Grimmett2018MNRAS,Ebinger2020ApJ}.
We translate the explosion energy into a minimum gas mass using:
\begin{equation}
    M_{\rm dil,min} \approx 1.9 \times 10^4 \msun\,E_{51}^{0.96} n_0^{-0.11}
\end{equation}
from \citet{Magg2020}. Though this limit was derived assuming spherical symmetry and a homogeneous ISM, it was validated by cosmological radiation hydrodynamic simulations \citep[][]{Magg2022}.
$E_{\rm 51}$ is the kinetic energy in units of $10^{51}$ erg (or 1 B) and $n_0$ is the ISM density in units of cm$^{-3}$.
Assuming a hydrogen mass fraction $X=0.75$ and A(Fe)$_\odot=7.50$, the \emph{maximum} metallicity achievable by a given supernova model and ISM density is given by:
\begin{equation}
    \mbox{[Fe/H]} < -2.40 + \log\frac{M_{\rm Fe}}{0.1 \msun} - 0.96 \log E_{51} + 0.11 \log n_0
\end{equation}
This calculation assumes a homogeneous ISM, but inhomogeneous mixing tends to exacerbate the problem, as the denser gas that turns into stars is more resistant to metal pollution \citep{Magg2020,Magg2022}.

\begin{figure*}
    \centering
    \includegraphics[width=\linewidth]{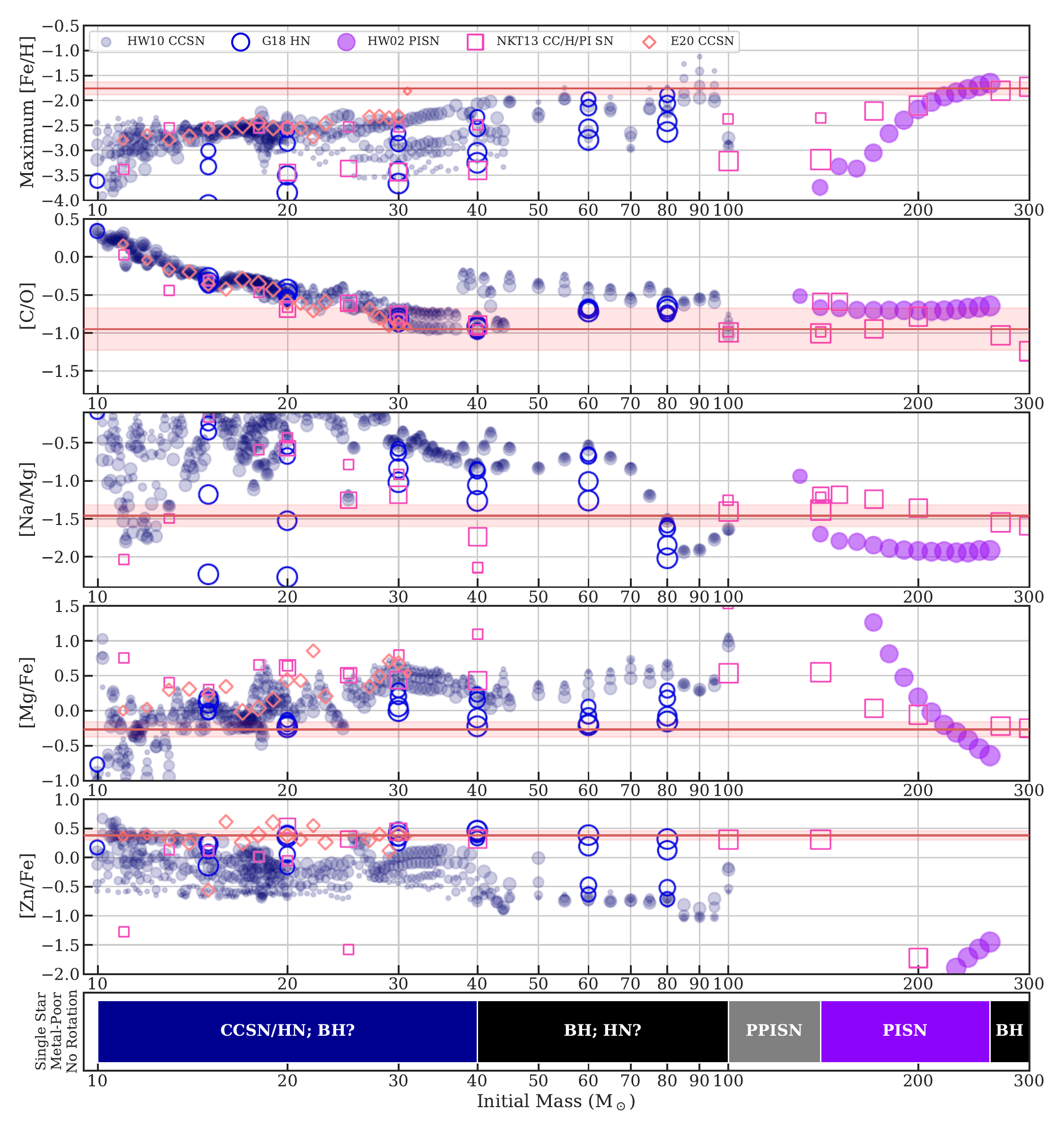}
    \caption{
    Initial mass vs maximum [Fe/H], [C/O], [Na/Mg], [Mg/Fe], and [Zn/Fe]
    for zero-metallicity supernovae from five yield grids: 
    CCSN from \citealt{Heger10} (HW10, fiducial mixing, $E = 0.3-2.4$B in steps of 2x); HN from \citealt{Grimmett2018MNRAS} (G18, no mixing, $E=5,10,50,100$B); PISN from \citealt{Heger2002ApJ} (HW02); all types of SN from \citealt{Nomoto2013} (NKT13), and CCSN from \citealt{Ebinger2020ApJ} (E20).
    Point sizes are proportional to log explosion energy (0.3-100B).
    The horizontal red lines and shaded regions show the abundance of \TheStar.
    Top row: maximum [Fe/H] strongly prefers higher mass progenitor stars with $M > 50 \msun$. Larger explosion energies tend to synthesize more Fe but dilute into more H, with the balance indicated by vertical trends in point sizes.
    Second row: [C/O] rules out low-mass CCSN ($M \lesssim 20 \msun$).
    Third and fourth row: [Na/Mg] and [Mg/Fe] disfavor intermediate-mass CCSN ($20-80 \msun$).
    Higher energy HN match the abundances better and allow progenitors down to $40 \msun$, but these all strongly violate the [Fe/H] constraint.
    E20 is removed from the [Na/Mg] plot as the Na yields are not predicted.
    Fifth row: PISNe ($M > 140 \msun$) are unable to produce significant Zn.
    Bottom row: estimated stellar fate given initial mass for a single, metal-poor, non-rotating star \citep{Heger2003}.
    }
    \label{fig:abund_Mini}
\end{figure*}

We plot this maximum [Fe/H] with $n_0=1$ for several nucleosynthesis models in the top panel of Figure~\ref{fig:abund_Mini}.
The [Fe/H] of \TheStar is shown by a red shaded band, and it can only be achieved in extreme explosions of massive stars:
either progenitor stars with $M > 50 \msun$, or pair instability supernovae with initial mass $M \gtrsim 200 \msun$.
Less massive progenitors simply do not produce enough iron or dilute into a small enough hydrogen mass to explain the metallicity of \TheStar.
The one exception is an engine-driven supernova by \citet{Ebinger2020ApJ}, which has $M_{\rm Fe} \sim 0.1\msun$ and energy $E \approx 0.3$B. These models are the most self-consistent CCSN explosions shown, but they do not include any fallback from a reverse shock, which is estimated to be about 0.1 \msun \citep{Perego2015ApJ} and would substantially lower the maximum [Fe/H].
The other core-collapse supernova and hypernova models are exploded with parameterized models, where the explosion energy and mixing/fallback are varied freely or fixed to reproduce certain observations.
At a fixed progenitor, higher explosion energies eject more Fe but dilute into more gas.\footnote{This figure shows only a selection of zero-metallicity supernovae for clarity, but all models in these and other grids are shown in Appendix~\ref{app:nuc}.}

\subsection{Nucleosynthetic Origin}\label{sec:nucleosynthesis}

\begin{figure}
    \centering
    \includegraphics[width=\linewidth]{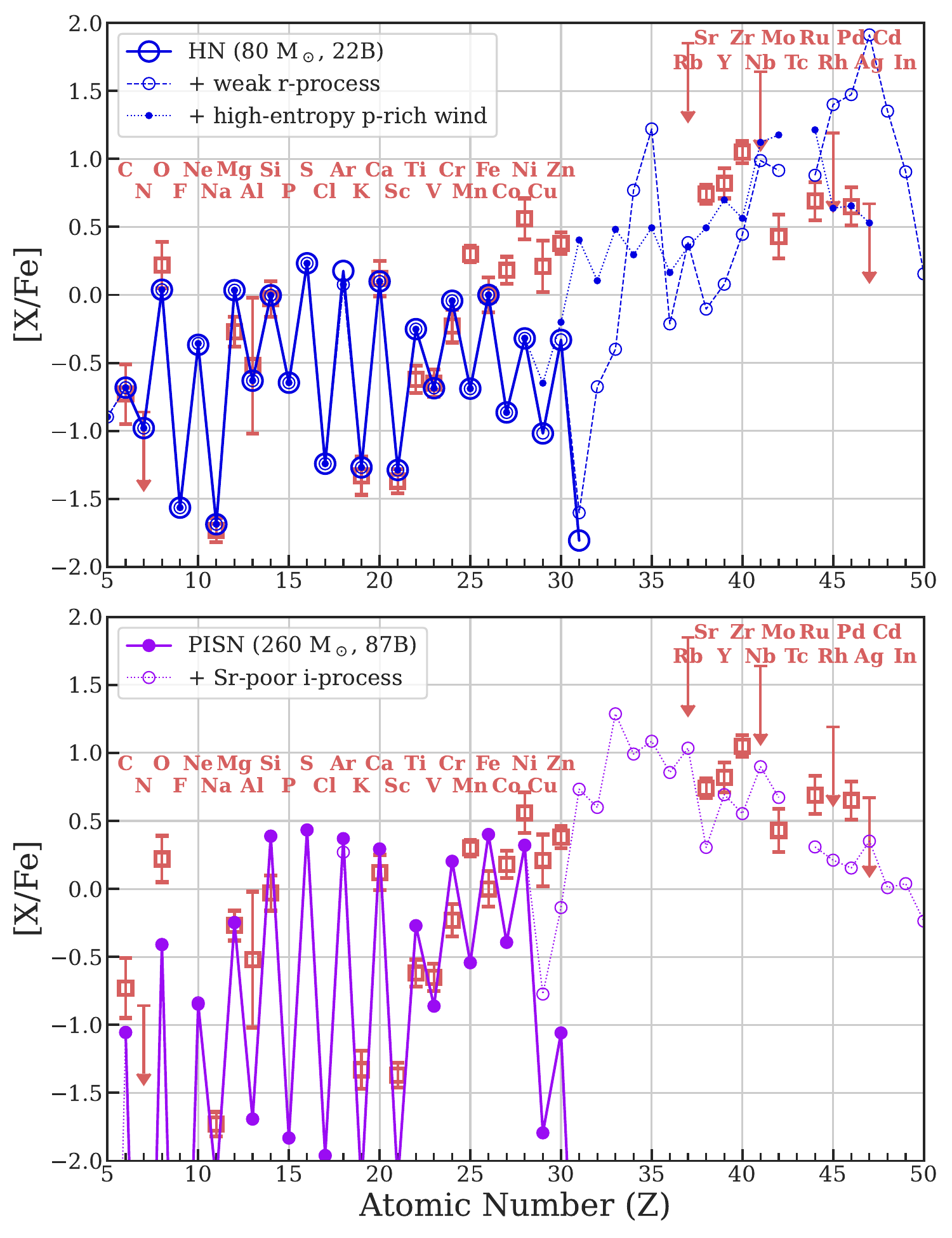}
    \caption{
    Representative best matching nucleosynthesis models.
    Abundances of \TheStar are given as red squares with error bars.
    In the top panel, the large open blue circles connected by a solid line indicate the best-fit HN model \citep{Grimmett2018MNRAS}
    The dotted blue line with small points and dashed blue line with small open circles indicate adding the HN to a high-entropy proton-rich wind pattern \citep{Bliss2018ApJ} and weak r-process pattern \citep{Holmbeck2023}, respectively.
    In the bottom panel, the solid purple circles with solid line indicate the best-fit PISN model \citep{Heger2002ApJ}.
    The dotted purple line with small open circles indicates the same model with a modified i-process (calculated using the framework in \citealt{Roederer2022}, see Appendix~\ref{app:nuc} for details).
    The HN is the best fit to the lighter elements and has leeway to fit the heavier elements, but the high energy underpredicts [Fe/H].
    The PISN easily matches the high [Fe/H], and we speculate that a full calculation of a metal-enriched PISN progenitor and explosion with i-process could remedy the disadvantages seen here.
    }
    \label{fig:nucl_best}
\end{figure}

We searched through several grids of nucleosynthesis models to determine what type of supernova could explain the abundance pattern of \TheStar. 
Here, we discuss the key element ratios that distinguish between progenitors, as illustrated in Figure~\ref{fig:abund_Mini}.
We primarily examine zero-metallicity supernova progenitors, as they have the largest range of model predictions, as well as lower neutron fractions that naturally result in a strong odd-even effect. However, we find no reason to exclude progenitors up to the ISM constraint $\mbox{[Fe/H]} \lesssim -3$ (Fig~\ref{fig:fehxh}).
Overall, the best models all invoke $M \gtrsim 50 \msun$ progenitors, and the best match out of current models is achieved by metal-free $80 \msun$ hypernovae (Fig~\ref{fig:nucl_best}).
However, we were unable to find any model that could explain all abundance features.
A detailed discussion is given in Appendix~\ref{app:nuc} for nucleosynthesis experts.

We first reject electron-capture supernovae ($M \sim 8-10 \msun$, \citealt{Doherty2017,Wanajo2018ApJ}), pulsational pair-instability supernovae (PPISNe, $100 \lesssim M/\msun \lesssim 140$, \citealt{Woosley2017}), and general relativistic instability supernovae ($M > 10^4 \msun$, \citealt{Chen2014ApJ}) as possibilities. These all produce extremely high ratios of [C, N, O/Fe] inconsistent with \TheStar.

We next examine deaths of $10-100 \msun$ stars. We split these models into lower energy core-collapse supernovae (CCSNe) powered by the ordinary neutrino-driven mechanism ($E \lesssim 2$B); and higher energy hypernovae (HNe), which likely require extra energy from rotation and jets, perhaps driven by black hole accretion disks or millisecond magnetars \citep[see references in][]{Grimmett2021MNRAS}.
It is broadly expected that $10-40\msun$ stars can explode as both CCSNe and HNe, while $40-100\msun$ stars probably need extra energy from a HNe to explode if they do at all \citep[][bottom row of Fig~\ref{fig:abund_Mini}]{Heger2003}\footnote{It is well-known that the explosion landscape from $10-100\msun$ is not monotonic in initial mass, instead showing islands of explodability \citep[e.g.,][]{Pejcha2015,Sukhbold2016,Burrows2018,Boccioli2023}. However, due to reduced mass loss in metal-poor stars, current self-consistent models of neutrino-driven explosions have all stars above 40\msun collapsing to black holes \citep{Ebinger2020ApJ}.}.
The most common SNe with $M \lesssim 20 \msun$ can be rejected due to their higher C/O yield ratios, a robust prediction of stellar evolution \citep[][second row of Fig~\ref{fig:abund_Mini}]{Ishigaki2018}. 
Above $20\msun$, CCSNe eject large amounts of hydrostatically synthesized elements like O and Mg, resulting in high [Na/Mg] and [Mg/Fe] ratios that conflict with \TheStar (third and fourth rows of Fig~\ref{fig:abund_Mini}). However, higher energy explosions can reduce the [Na/Mg] and [$\alpha$/Fe] yields to be consistent with \TheStar (see HN models 
 from \citealt{Grimmett2018MNRAS} in Fig~\ref{fig:abund_Mini}).
Both CCSNe and HNe can likely synthesize the light neutron-capture elements seen in \TheStar, either through neutrino-driven winds \citep{Frohlich2006PRL,Bliss2018ApJ,Wanajo2018ApJ} or accretion disk winds \citep{Pruet2004,Surman2006,Siegel2019}, though it is hard to reproduce the exact pattern observed.

Overall, $40-100 \msun$ HNe can broadly match most major abundance features of \TheStar.
The best match we found is a 80\msun 22B model \citep{Grimmett2018MNRAS} shown in Fig~\ref{fig:nucl_best}.
However, we emphasize that the full nucleosynthetic pattern is not fit by \emph{any} existing model.
In particular, the Fe peak abundances are extremely difficult to explain: models must simultaneously produce low [Sc,Ti,V/Fe] (associated with lower energy) and high [Ni,Zn/Fe] (associated with higher energy); and high [Mn/Fe] (associated with high neutron fractions) but low odd-even ratios (associated with low neutron fractions).
The higher energy HNe with $E \gtrsim 10$B that give the best matches to the nucleosynthesis pattern predict maximum metallicities $\mbox{[Fe/H]} \ll -2$, strongly violating the metallicity constraint.
Additionally, the low N and Ba restrict the progenitor's rotation, as rotational mixing increases N and would over-produce Ba through the s-process if there are seed nuclei \citep{Ekstrom2008AA,Pignatari08,Frischknecht16,Choplin2018AA}.
This motivates considering other sources.

One intriguing possibility is that \TheStar was enriched by a pair instability supernova (PISN). Metal-free high mass PISNe with $M \gtrsim 200 \msun$ produce a very large amount of Fe, a strong odd-even effect, and low [$\alpha$/Fe] ratios \citep{Heger2002ApJ,Takahashi2018}. These ``smoking gun'' signatures of PISNe qualitatively match \TheStar. However, all existing PISNe models produce negligible $\mbox{[Zn/Fe]} < -1$ and no neutron-capture elements \citep[][5th row of Fig~\ref{fig:abund_Mini}]{Heger2002ApJ,Salvadori2019}. Thus, standard PISNe are unable to explain \TheStar's abundance pattern.
Still, the strong association with PISN motivated some additional exploration, and we found that the intermediate neutron-capture (i-process) nucleosynthesis in PISN progenitors could be a promising mechanism. The i-process can occur if convection causes protons to be ingested into a He shell \citep{Herwig2014,Woodward2015,Roederer2016c,Clarkson18,Banerjee18a}, which generates neutrons that capture onto seed nuclei. With the right neutron exposure and initial composition (see Appendix~\ref{app:nuc}), the i-process converts Fe into enhanced Zn and Sr-Pd without significant Ba, which qualitatively matches \TheStar. This explanation would require a metal-enriched PISN progenitor, implying that luminous PISNe could be found at later times than usually assumed \citep{Hartwig2018b}. We show this speculative model in Figure~\ref{fig:nucl_best}, which simply adds the i-process pattern to a PISN yield.

\section{Conclusion} \label{sec:conclusion}

We have presented the extreme chemical abundance pattern of the star \TheStar (Figure~\ref{fig:xfe_saga}).
The low abundances of Na, K, Sc, and Ba and the high abundances of Fe peak elements and Sr-Pd show that most of the metals in this star came from a single nucleosynthetic source (Figure~\ref{fig:fehxh}).
The high overall metallicity, low [C/O] ratio, and strong odd-even effect together combine to prefer progenitors with mass $>50\msun$ (Figure~\ref{fig:abund_Mini}).
However, the detailed abundance pattern, especially in the iron peak, is not fully explained by any existing models of nucleosynthesis in massive stars (Figure~\ref{fig:nucl_best}).
One possibility is the source might be a hypernova with progenitor mass $\sim 80 \msun$, which would be the first example of an early supernova from a star with initial mass between $50-100\msun$. Alternatively, the star might indicate i-process nucleosynthesis in the progenitor of a metal-enriched pair instability supernova, which would be the first example of a metal-enriched PISN. There may be other pathways that we did not consider.

\TheStar shows that current models of massive, metal-poor star nucleosynthesis are still quite limited, challenging the next generation of models. We suggest that \TheStar points to the inherent multi-dimensional nature of nucleosynthesis in massive stars, such as convective nuclear burning that likely impacts the iron peak yields \citep[e.g.,][]{Herwig2014,Woodward2015,Curtis19,Fields2020,Burrows2021,Sieverding2023}.
We highlight rotation and jets for massive stars $\sim 80 \msun$ and i-process from proton ingestion in massive PISN progenitors as fruitful paths for exploration.
Another important consideration is binarity, as essentially all massive stars are in binaries, and most in interacting binaries \citep{Sana2012}. There are still no studies of nucleosynthesis in metal-poor or metal-free supernovae of interacting binaries. This may be important especially because metal-poor stars don't lose much mass through winds, but they can lose significant mass through binary interactions \citep{deMink2008}.
To our knowledge there also have been no studies of nucleosynthesis from interacting stars sufficiently massive for a PISN, and it would be interesting to see if interactions could produce black hole or neutron star remnants that would remedy the deficiencies of single PISN models.
Finally, we note that whatever produces the abundance signature of \TheStar is probably very rare, otherwise this pattern would probably have already been previously discovered in the thousands of existing metal-poor stellar abundance data \citep{Suda2008,jinabase,Li2022}. Converting this frequency to a volumetric rate estimate would require a model of dwarf galaxy and stellar halo formation, which is outside of the scope of this Letter.

Though we focused here on the nucleosynthetic implications, we speculate that \TheStar's unique composition implies that the rare supernova events that could explain this signature should also be found in upcoming large transient surveys, such as Rubin/LSST \citep{LSSTScienceCollaboration2009}.
If the chemical signature is due to an unusual PISN, the heavy Fe peak and neutron-capture elements point to the presence of a neutron star or black hole remnant involved in the explosion, which may result in unusual observational features of slowly evolving superluminous supernovae \citep{GalYam2019,Nicholl2021}.
If the chemical signature is instead due to a massive CCSN or HN, there should be supernovae of massive stars with low kinetic energy but relatively high $^{56}$Ni luminosity \citep[e.g., SN2008ha][]{Foley2010,Moriya2010}.
Finally, if binarity is needed, it is possible that \TheStar has implications for features or outliers in the compact binary merger mass spectrum \citep{GWTC3Pops,Farah2023}.

\TheStar was identified in the first year of SDSS-V observing and after only one semester of followup of the metal-poor and low-$\alpha$ stars.
The rapid discovery suggests that many more rare nucleosynthesis events like this should be found in the current and upcoming era of large spectroscopic surveys, and \TheStar emphasizes the importance of searching in multiple abundance dimensions rather than just at low metallicities.
The unique chemical signature of \TheStar would also make it very easy to chemically identify companions from any accreted kinematic group. We searched for stars in APOGEE DR17 with similar kinematics and low Mg abundances from \citet{Horta2023} and examined all stars with low [Al/Mg] ratios that could potentially be analogues of this star. We were unable to find any candidates, but the current generation of large spectroscopic surveys may turn up future counterparts.

\section*{Acknowledgements}
We acknowledge extremely helpful comments from the referee, Stan Woosley, that significantly improved the readability and clarity of this paper.
This paper includes data gathered with the 6.5~meter Magellan Telescopes located at Las Campanas Observatory, Chile.
A.P.J. acknowledges the University of Chicago’s Research Computing Center for their support. This work benefited from a workshop supported by the National Science Foundation under Grant No. OISE-1927130 (IReNA), the Kavli Institute for Cosmological Physics, and the University of Chicago Data Science Institute.
A.P.J. acknowledges Ani Chiti and Guilherme Limberg for helpful conversations.
We acknowledge support awarded by the U.S. National Science Foundation (NSF) grants:
AST-2206264 (A.P.J., P.T., S.A.U.),
OISE-1927130 (IReNA; A.P.J., M.P.),
PHY-1430152 (JINA-CEE; A.P.J., M.P.),
AST-2303869 (AAPF; S.C.),
and AST-2202135 (AAPF; E.J.G.).
We acknowledge funding from the European Research Council (ERC) under several of the European Union's Horizon 2020 research and innovation programmes: Grant agreement Nos. 
101008324 (ChETEC-INFRA; M.P.), 
949173 (M.B.), 
852839 (J.A., C.L.), 
and 833925 (STAREX; G.M.).
A.H. was supported, in part, by the Australian Research Council (ARC) Centre of Excellence (CoE) for All Sky Astrophysics in 3 Dimensions (ASTRO 3D), through project number CE170100013 and acknowledges software development support from Astronomy Australia Limited's ADACS scheme (Project IDs AHeger\_2022B, AHeger\_2023A) for the development of the StarFit code used here.
M.P. thanks the support from the NKFI via K-project 138031 and the ERC Consolidator Grant (Hungary) programme (RADIOSTAR, G.A. n. 724560). M.P. acknowledges the support to NuGrid from STFC (through the University of Hull’s Consolidated Grant ST/R000840/1), and ongoing access to {\tt viper}, the University of Hull High Performance Computing Facility. M.P. acknowledges the support from the "Lend{\"u}let-2023" LP2023-10 Programme of the Hungarian Academy of Sciences (Hungary).
M.B. is supported through the Lise Meitner grant from the Max Planck Society. We acknowledge support by the Collaborative Research centre SFB 881 (projects A5, A10), Heidelberg University, of the Deutsche Forschungsgemeinschaft (DFG, German Research Foundation). 
H. R. acknowledges support from a Carnegie Fellowship.
C.F. was supported by United States Department of Energy, Office of Science, Office of Nuclear Physics (award number DE-FG02-02ER41216).

Funding for the Sloan Digital Sky Survey V has been provided by the Alfred P. Sloan Foundation, the Heising-Simons Foundation, the National Science Foundation, and the Participating Institutions. SDSS acknowledges support and resources from the Center for High-Performance Computing at the University of Utah. The SDSS web site is \url{www.sdss.org}.

SDSS is managed by the Astrophysical Research Consortium for the Participating Institutions of the SDSS Collaboration, including the Carnegie Institution for Science, Chilean National Time Allocation Committee (CNTAC) ratified researchers, the Gotham Participation Group, Harvard University, Heidelberg University, The Johns Hopkins University, L'Ecole polytechnique f\'{e}d\'{e}rale de Lausanne (EPFL), Leibniz-Institut f\"{u}r Astrophysik Potsdam (AIP), Max-Planck-Institut f\"{u}r Astronomie (MPIA Heidelberg), Max-Planck-Institut f\"{u}r Extraterrestrische Physik (MPE), Nanjing University, National Astronomical Observatories of China (NAOC), New Mexico State University, The Ohio State University, Pennsylvania State University, Smithsonian Astrophysical Observatory, Space Telescope Science Institute (STScI), the Stellar Astrophysics Participation Group, Universidad Nacional Aut\'{o}noma de M\'{e}xico, University of Arizona, University of Colorado Boulder, University of Illinois at Urbana-Champaign, University of Toronto, University of Utah, University of Virginia, Yale University, and Yunnan University.

This work has made use of data from the European Space Agency (ESA) mission
{\it Gaia} (\url{https://www.cosmos.esa.int/gaia}), processed by the {\it Gaia}
Data Processing and Analysis Consortium (DPAC,
\url{https://www.cosmos.esa.int/web/gaia/dpac/consortium}). Funding for the DPAC
has been provided by national institutions, in particular the institutions
participating in the {\it Gaia} Multilateral Agreement.

This research has made use of NASA’s Astrophysics Data System Bibliographic Services; the arXiv preprint server operated by Cornell University; and the SIMBAD databases hosted by the Strasbourg Astronomical Data Center.
This research has made use of the Keck Observatory Archive (KOA), which is operated by the W. M. Keck Observatory and the NASA Exoplanet Science Institute (NExScI), under contract with the National Aeronautics and Space Administration.


%

\vspace{5mm}
\facilities{Sloan, Magellan (MIKE), Gaia}


\software{
\texttt{MOOG} \citep{Sneden73,Sobeck11},
\texttt{Turbospectrum} \citep{Plez2012},
\texttt{TSFitPy} \citep{Gerber2023},
\texttt{smhr} \citep{Casey14,Ji2020},
\texttt{emcee} \citep{emcee},
\texttt{gala} \citep{gala,adrian_price_whelan_2022_7299506},
\texttt{minesweeper} \citep{Cargile2020},
\texttt{dynesty} \citep{Speagle2020},
\texttt{isochrones} \citep{mor15},
\texttt{MultiNest} \citep{fer08,fer09,fer19},
\texttt{R} \citep{r23},
\texttt{StarFit} \citep{Heger10},
\texttt{numpy} \citep{numpy}, 
\texttt{scipy} \citep{scipy}, 
\texttt{matplotlib} \citep{matplotlib},
\texttt{pandas} \citep{pandas},
\texttt{seaborn} \citep{seaborn},
and \texttt{astropy} \citep{astropy,astropy2}
}



\appendix

\section{Observational Analysis Details}\label{app:sp}

\subsection{BOSS/MINESweeper Analysis}

The SDSS-V BOSS \citep{Gunn2006,Smee2013} spectra of \TheStar (\GaiaID, $\ell,b=233.178248,+35.189507$) were observed on 2022~April~25, reduced using the BOSS data reduction pipeline (\citealt{Bolton2012,Dawson2013}, Morrison et al. in prep), and analyzed using \minesweeper{} \citep{Cargile2020}, which performs a spectrophotometric fit simultaneously to the BOSS spectrum, the Gaia DR3 parallax \citep{GaiaCollaboration2021}, and all available broadband photometry \citep{Gunn1998, skr06, Mainzer2014, Chambers2016}. 
Stars are constrained to lie on MIST isochrones \citep{Choi2016}, and stellar parameters are sampled using the \texttt{dynesty} nested sampling code \citep{Speagle2020}. 
The spectroscopic fit was restricted to the region around Mg b (4750-5550\,\AA{}), as this is the region where the spectral models have been well-calibrated for the H3 Survey \citep{Conroy2019}. 
\minesweeper{} provides the effective temperature, surface gravity, bulk metallicity, and [$\alpha$/Fe] abundance based on this fit, where the $\alpha$ is primarily determined by the Mg b lines. 
The \minesweeper{} parameters and formal uncertainties for \TheStar were $\teff=5220 \pm 30$K, $\logg=2.57 \pm 0.06$, $\feh=-1.9 \pm 0.1$, and $\alphafe = 0.03 \pm 0.17$.

\begin{figure}
    \centering
    \includegraphics[width=0.8\linewidth]{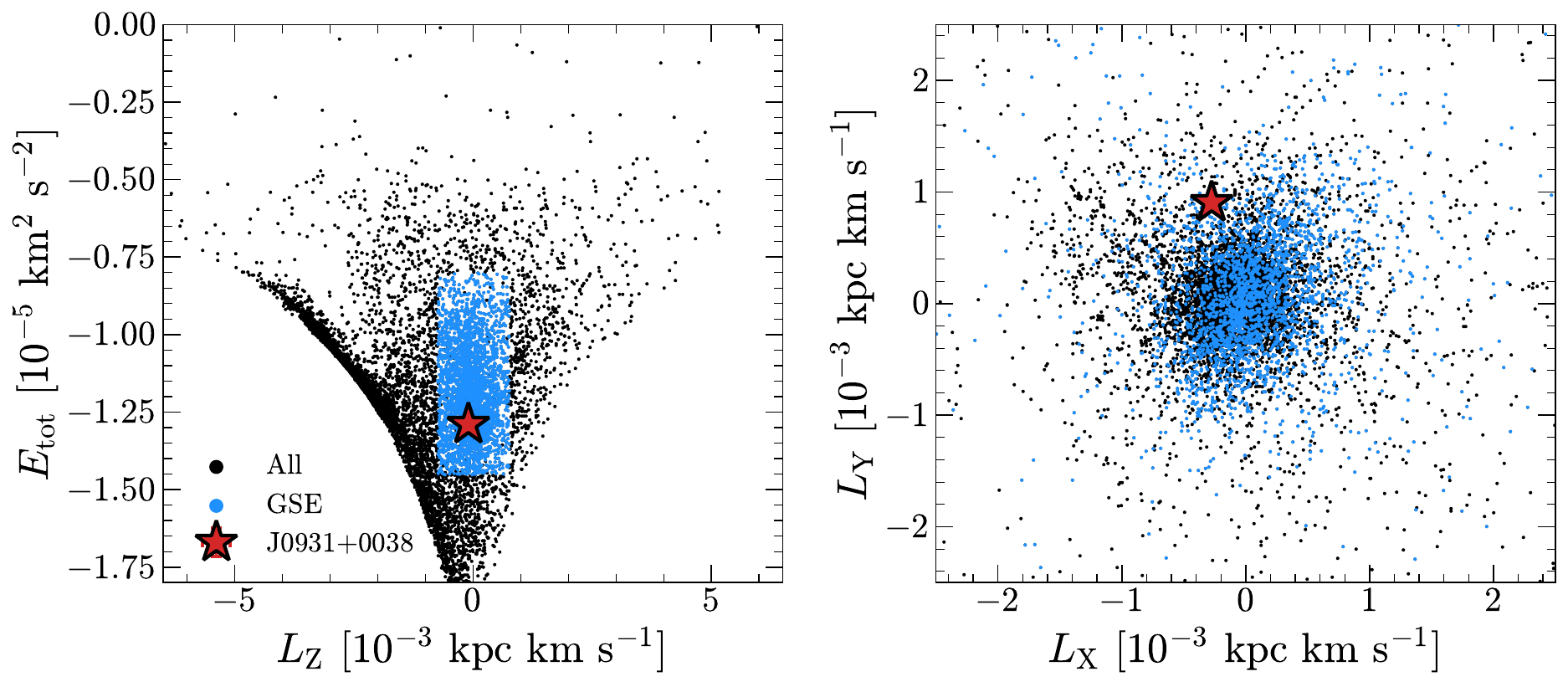}
    \caption{Kinematics of \TheStar (red star) compared to red giants observed by the SDSS-V halo program. Stars plausibly belonging to the Gaia Sausage Enceladus (GSE) merger are highlighted in blue.}
    \label{fig:LxLy}
\end{figure}

The uniform \minesweeper{} analysis of all SDSS-V halo targets provides their 3D positions and velocities and enables investigation of their kinematics.
The total specific energy was calculated using the latest \texttt{MilkyWayPotential2022} in \texttt{gala} \citep{gala, adrian_price_whelan_2022_7299506}, which matches the rotation curve data from \cite{Eilers2019}. 
The energy and three components of angular momentum of \TheStar are shown in Figure~\ref{fig:LxLy} as a large red star, compared to all other stars observed by the SDSS-V halo cartons from Internal Product Launch 2 as small black points.
The energy and $L_Z$ clearly show \TheStar is a halo star, with an eccentric radial orbit ($e=0.84$, pericenter=1.1kpc) consistent with the Gaia-Sausage-Enceladus (GSE) dwarf galaxy merger \citep{Belokurov2018a,Helmi2018}.
However, cosmological simulations suggest that this region of kinematic space is crowded, so the majority of metal-poor stars on GSE-like orbits actually do not come from GSE \citep{Brauer2022, Orkney2023}.
Tailored GSE-like merger models also show that its debris is confined to a region $|L_x|,|L_y| \lesssim 800 \unit{kpc} \unit{km} \unit{s}^{-1}$ \citep{Naidu2021,Amarante2022}.
Indeed \TheStar lies outside the bulk of GSE stars in $L_X-L_Y$ space, which emphasizes the importance of checking multiple kinematic quantities when correlating halo structures.
\TheStar is thus likely accreted as part of a now-disrupted dwarf galaxy, but probably not the large GSE merger itself.

\subsection{Stellar Parameters}
Our analysis primarily used the 1D ATLAS model atmospheres \citep{Castelli04} and the MOOG radiative transfer code \citep{Sneden73} including scattering \citep{Sobeck11} and assuming local thermodynamic equilibrium (LTE).
The abundance analysis was conducted in the \code{smhr} environment \citep{Casey14}.
The line lists and atomic data were selected from a combination of lines from \citet{Roederer2018} and \citet{Ji2020}, with atomic data adopted from linemake \citep{Placco2021}.
Stellar parameters were determined using a combination of spectroscopy, photometry, isochrones, and Gaia parallax.
We adopted an effective temperature of 5200K based on the \minesweeper{} spectrophotometric results, then determined other parameters spectroscopically.
The surface gravity required to balance the neutral and ionized iron abundances was $\logg = 2.75$, the microturbulence required to balance the Fe II line abundances was 1.65 km/s, with a model metallicity of -1.9 and using solar-scaled abundances for the ATLAS atmosphere composition ($\alphafe=0$).

Stellar parameters and uncertainties were checked using two independent fits to the spectral energy distribution and parallax.
First, we performed an analysis of the broadband spectral energy distribution (SED) of the star together with the {\it Gaia\/} DR3 parallax \citep[with a systematic offset applied; see, e.g.,][]{StassunTorres:2021} following the procedures described in \citet{Stassun:2016,Stassun:2017,Stassun:2018}. We pulled the $JHK_S$ magnitudes from {\it 2MASS}, the W1--W3 magnitudes from {\it WISE}, the $G_{\rm BP}$ and $G_{\rm RP}$ magnitudes from {\it Gaia}, the $grizy$ magnitudes from {\it Pan-STARRS}, and the NUV magnitude from {\it GALEX}. We also used the {\it Gaia\/} spectrophotometry spanning 0.4—1.0~$\mu$m. Altogether, the available photometry spans the full stellar SED over the wavelength range 0.2--10~$\mu$m. The {\it GALEX\/} flux in particular helps to constrain the metallicity and the {\it Gaia\/} spectrophotometry provides an especially strong constraint on the overall absolute flux calibration. 
We then performed a fit using PHOENIX stellar atmosphere models \citep{Husser:2013}, with the free parameters being the effective temperature ($T_{\rm eff}$) and metallicity ([Fe/H]), as well as the extinction, which we set to the maximum line-of-sight value $A_V = 0.15 \pm 0.02$ from the Galactic dust maps of \citet{Schlegel:1998} due to the system's large distance. We initially assumed a surface gravity $\log g \approx 2.5$ given the likely evolutionary state of the star. The resulting fit has a best-fit $T_{\rm eff} = 5250 \pm 50$~K and [Fe/H] = $-2.3 \pm 0.3$. Integrating the (unreddened) model SED gives the bolometric flux at Earth, $F_{\rm bol} = 7.988 \pm 0.092 \times 10^{-11}$ erg~s$^{-1}$~cm$^{-2}$. Taking the $F_{\rm bol}$ and $T_{\rm eff}$ together with the {\it Gaia\/} parallax, gives the stellar radius, $R_\star = 8.65 \pm 0.9 R_\odot$.
In addition, we estimate the stellar mass to be $M_\star = 0.8 \pm 0.1 \msun$, as \TheStar is a metal-poor old red giant.
The mass and radius together confirm $\logg \approx 2.5$.

Second, we also derived the stellar parameters
of \TheStar using the \texttt{isochrones}
\citep{mor15} package to execute with \texttt{MultiNest}
\citep{fer08,fer09,fer19} a simultaneous Bayesian fit of the MIST
isochrone grid \citep{pax11,pax13,pax18,pax19,jer23,dot16,Choi2016} to a
curated collection of data for the star.  We fit (1) Galaxy Evolution
Explorer (GALEX) GUVcat\_AIS NUV \citep{bia17}, Sloan Digital Sky Survey
(SDSS) Data Release (DR) 18 $ugiz$ \citep{fuk96,gun98,yor00,doi10,Abdurrouf2022},
Gaia DR2 $G$ \citep{gai16,gai18,are18,eva18,rie18}, Two-micron
All-sky Survey (2MASS) $JHK_{\text{s}}$ \citep{skr06},
and Wide-field Infrared Survey Explorer CatWISE2020 $W1W2$
photometry \citep{wri10,mar21}, (2) a zero point-corrected Gaia DR3
parallax \citep{gai21,fab21,lin21a,lin21b,row21,tor21}, and (3) an
estimated reddening value based on a three-dimensional reddening map
\citep{gre14,gre19}.  We use a log uniform age prior between 8.0 Gyr and
13.7 Gyr, a uniform reddening prior between the estimated reddening value
minus/plus five times its uncertainty, and a distance prior proportional
to volume between the \citet{bai21} geometric distance minus/plus five
times its uncertainty.  We find the photospheric stellar parameters
$T_{\text{eff}} = 5140_{-10}^{+20}$ K, $\log{g} = 2.51_{-0.07}^{+0.07}$,
and $[\text{Fe/H}] = -1.63_{-0.01}^{+0.07}$.

As discussed in the main text, we adopted stellar parameters of $\teff = 5200 \pm 100$K, $\logg = 2.75 \pm 0.20$,  $\nu_t = 1.65 \pm 0.3 \kms$, $[M/H]=-1.9 \pm 0.1$, and $\alphafe=0.0$.
The temperature uncertainty of 100K resulted in correlated \logg and metallicity offsets of 0.20 dex and 0.1 dex.
The microturbulence uncertainty of 0.3 km/s is very conservative and represents the most extreme values found during line strength balance both in LTE and non-LTE in all permutations.
These were later propagated into all abundance uncertainties.

\subsection{LTE Abundance Analysis}
Chemical abundances were determined in 1D LTE using MOOG and ATLAS in \code{smhr}, with a mix of equivalent widths for isolated, unblended lines and syntheses for molecular bands, lines with hyperfine structure, or moderately blended lines.
The local continuum and smoothing were allowed to vary for each feature.
Syntheses were fit by minimizing a chi-square statistic (for more details, see \citealt{Ji2020}).
The final chemical abundance for each species was found as the unweighted average of individual line abundances.
The Fe in [X/Fe] refers to Fe I.
For non-detections, we synthesize a best-fit spectrum with no line, then calculate a formal $5\sigma$ upper limit by increasing the abundance until the $\chi^2$ changes by $5^2$. This assumes no uncertainties in the continuum, which is a good assumption given the high S/N of our spectrum, except for the N-H molecule where we visually estimated a very conservative upper limit for N-H (formally $>10\sigma$) with a synthetic spectrum.

Systematic abundance uncertainties due to stellar parameters were found by redetermining the chemical abundances at two alternate stellar parameter values based on the stellar parameter uncertainty:
($\teff$, $\logg$, $\nu_t$, [M/H]) = (5100\,K, 2.55, 1.65, $-2.0$) and (5200\,K, 2.75, 1.95, $-1.9$).
For Fe I, we sum the total difference in [Fe I/H] in quadrature for these two sets of stellar parameters and adopt that as the stellar parameter uncertainty on the absolute metallicity of the star.
For species other than Fe I, we adopt the difference in [X/Fe\,I] for each of these variations as the stellar parameter uncertainty and sum them in quadrature.
The latter accounts for the fact that [X/H] and [Fe/H] are highly correlated with respect to stellar parameters, so the relative abundance uncertainty is smaller (which is the relevant uncertainty when considering the total abundance pattern).

To investigate the systematic effect of our model atmosphere, line list, and radiative transfer code, we also analyzed a subset of the lines using 1D spherical MARCS model atmospheres \citep{Gustafsson2008}, the most recent version of \code{Turbospectrum} in \code{TSFitPy} \citep{Plez2012,Gerber2023}, and a linelist from Gaia-ESO \citep{Heiter2021} with gaps filled and the range further extended with VALD \citep{VALD}. 
We adopt the MARCS model atmospheres, and the analysis used slightly different stellar parameters: $\teff=5200\,$K, $\logg=2.60$, $\nu_t=1.6\kms$, $\feh=-1.85$, and $\alphafe=+0.4$ (due to standard MARCS grid).
The final LTE abundance differences are all within 0.1 dex, with the exception of aluminum which will be discussed later. We thus decided to adopt a minimum 0.1 dex systematic uncertainty per line, such that the total systematic uncertainty goes down as the square root of the number of lines.

In summary, the adopted abundance uncertainty is the quadrature sum of four components: the line-to-line standard deviation for a given species, a minimum systematic of 0.1 dex divided by the square root of the number of lines per element, the stellar parameter error after changing $\teff$ and $\logg$ with their correlated uncertainties, and the stellar parameter error after changing $\nu_t$ by 0.3 \kms.

\subsection{Non-LTE Corrections}

Non-LTE (NLTE) corrections for most elements were determined using \code{TSFitPy} \citep{Gerber2023}.
We fit the same lines used for the MOOG/ATLAS analysis, but restricted to the wavelength range 3700-9200{\AA} using the Gaia-ESO linelist with gaps filled and the range extended using the VALD line list. The NLTE corrections were determined using the standard concept of the NLTE abundance correction \citep{Bergemann2014}, which represents the difference in abundance that is required to match the equivalent width (EW) of a NLTE model line to that of the LTE line computed using the identical values of stellar parameters.
The elements and model atom references were:
Oxygen \citep{Bergemann2021},
Sodium \citep{Larsen2022},
Magnesium \citep{Bergemann2017},
Silicon \citep{Bergemann2013, EMagg2022},
Calcium \citep{Mashonkina2017, Semenova2020},
Titanium \citep{Bergemann2011},
Manganese \citep{Bergemann2019},
Iron \citep{Bergemann2012b, Semenova2020},
Cobalt \citep{Bergemann2010, Yakovleva2020},
Nickel \citep{Bergemann2021, Voronov2022},
Strontium \citep[][Gallagher et al. in prep.]{Bergemann2012a},
Yttrium \citep{Storm2023}, 
and Barium \citep{Gallagher2020}.
The average NLTE correction from all lines was taken as the total NLTE correction for that element.

A few elements are not currently included in \code{TSFitPy}, and we describe their NLTE corrections below.

\textit{Aluminum.}
Only the 3961{\AA} line is usable, and it is heavily blended with a strong Ca and H feature.
The abundance of Al is extremely uncertain as a result.
We measure [Al/Fe] in LTE through spectrum synthesis, and the LTE abundances from MOOG and \code{TSFitPy} differed by 0.2 dex, likely dominated by treatment of the blending Ca and H feature.
We then adopt the NLTE correction from the grid of \citet{Nordlander17} to get a $+0.65$ dex correction to [Al/Fe].
Preliminary calculations with other unpublished Al model atoms in \code{TSFitPy} suggested a smaller correction of $+0.33$ dex.
We thus adopt a very large uncertainty of 0.5 dex for the Al abundance to represent both the blending and NLTE correction uncertainty.

\textit{Potassium.}
The K abundance is derived from equivalent widths of both 7699{\AA} and 7665{\AA} line, which are both unaffected by telluric lines in this star.
Examining stars with similar stellar parameters in \citet{Reggiani19}, NLTE corrections for K from the 7699 line range from $-0.17$ to $-0.24$ dex.
We adopt a $-0.2$ dex correction, and increase the K uncertainty by adding 0.1 dex in quadrature.

\textit{Copper.}
We use the equivalent width of the 5105{\AA} line and adopt an empirical correction of $+0.35$ dex from \citet{RoedererBarklem2018}. 
This matches theoretical calculations \citep{Andrievsky2018,Korotin2018} and we increase the uncertainty by adding 0.15 dex in quadrature to reflect the scatter from the theoretical calculations.

\textit{Elements without corrections.}
\citet{Roederer2022} provide a detailed accounting of what corrections might be expected based on comparisons of neutral and ionized lines in a star HD222925 with $\mbox{[Fe/H]}=-1.5$ and $\teff=5640$K.
Based on this we do not expect significant NLTE corrections for zinc, zirconium, molybdenum or ruthenium; while possible NLTE corrections for rhodium and palladium are unconstrained.

\subsection{Evolutionary State Corrections}

\TheStar has $\logg=2.75$, so it has passed the first dredge up but not the red giant branch bump, and a small amount of C is converted to N.
To account for this difference, we examined metal-poor red giants in APOGEE DR17 with $\feh < -1.5$. Stars after the first dredge up have [C/N] higher by 0.2 dex, where [C/Fe] is lower by 0.1 dex and [N/Fe] is higher by 0.1 dex.
For \TheStar, we thus increase [C/H] by $+0.1$ dex, decrease the [N/H] upper limit by $-0.1$ dex, and increase each element's uncertainty by adding 0.2 dex in quadrature.

We also measured a Li abundance of A(Li)$=1.15 \pm 0.12$. Given the \logg of this star, this Li abundance is consistent with Li depletion in the first dredge up \citep[e.g.,][]{Tayar2022}, a good independent check on the stellar parameters.

\subsection{Binarity and Photometric Variability}
\TheStar displays no evidence for a present-day binary companion.
The heliocentric radial velocity for MIKE was found to be $105.5 \pm 0.4 \kms$ \citep[measured with the method in][]{Ji2020a}, while Gaia DR3 RVS reports $104.3 \pm 4.0 \kms$.
These velocities are consistent within uncertainties.
The velocity scatter from multiple Gaia RVS transits is large but typical for stars of similar spectral type, distance, and signal-to-noise \citep{Chance2022}, and there is no evidence for excess astrometric scatter \citep{Penoyre2020}.

Photometric variability could also be used to identify binary companions or measure solar-like oscillations.
\TheStar does not show up in \citet{Hon2021} as a solar-like oscillator.
We obtained the TESS light curve of \TheStar (TIC 383218318) using TESScut \citep{Brasseur2019} with a custom aperture and subtracting background flux. Following \citet{Avallone2022}, we normalized and smoothed each of the sectors and took a Fourier transform of the resulting light curve.  We do not see any evidence of periodic variability in the TESS light curve that would suggest detectable rotational modulation or oscillations, though \TheStar is relatively faint for TESS and has a limited time baseline.

\subsection{Comparison to Notable Stars}
\begin{figure}
    \centering
    \includegraphics[width=0.9\linewidth]{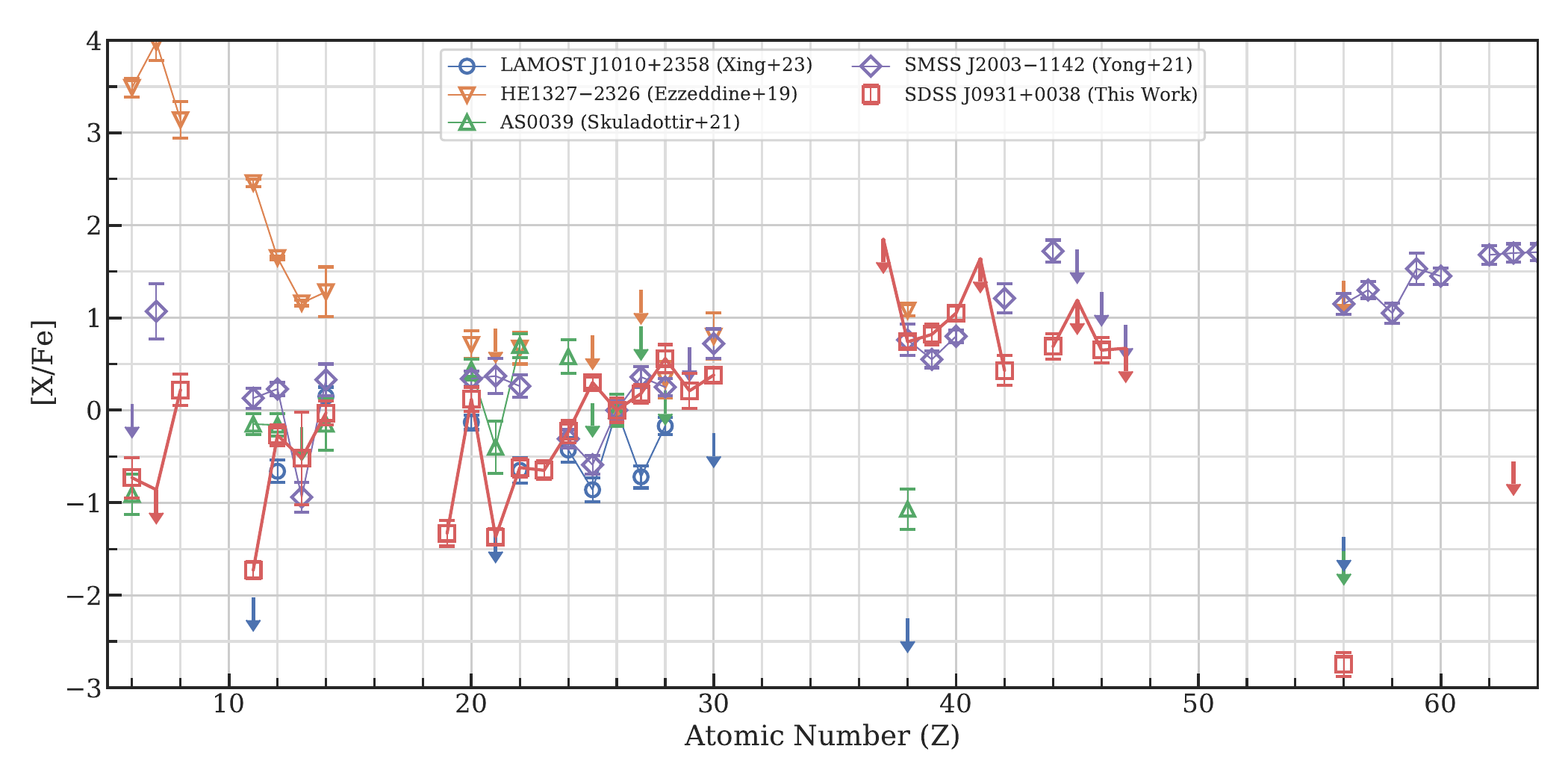}
    \caption{Chemical abundances of \TheStar compared to four other notable stars. LAMOST J1010$+$2358 has a pure signature of a pair instability supernova, while the other three are metal-poor stars whose compositions are currently best explained with hypernovae models.}
    \label{fig:litcomp}
\end{figure}

Figure~\ref{fig:litcomp} shows the [X/Fe] of \TheStar compared to four notable stars:
three hypernova candidate stars with varying neutron-capture element abundances 
HE1327$-$2326 \citep{Frebel2005,Ezzeddine2019}, 
SMSS~J2003$-$1142\citep{Yong2021}, 
and AS0039 \citep{Skuladottir2021};
and the pair instability supernova star J1010$+$2358 (\citealt{Xing2023}, though note it has recently been argued that this star is also consistent with an extreme CCSN, \citealt{Jeena2023}).

It is clear that \TheStar has an extreme abundance pattern even compared to these other notable stars. Because of its relatively high metallicity, it is also easier to measure many more elements.
The clear signature is that the odd elements Na, Al, K, and Sc ($Z=11,13,19,21$) in \TheStar are lower than almost all the other stars, with the exception of the PISN star J1010$+$2358.
The carbon and oxygen abundances ($Z=6,8$) are relatively low, in contrast with more metal-poor stars that tend to be carbon-enhanced like HE1327$-$2326 and SMSS J1605$-$1443.
Mn ($Z=25$) is unusually high, as nearly all metal-poor stars have [Mn/Fe] $<0$.
Co through Zn ($Z=27-30$) are also elevated, which is usually associated with hypernovae \citep{Ezzeddine2019,Yong2021}.

The three literature hypernova stars have very different neutron-capture patterns.
HE1327$-$2326 has high Sr ($Z=38$) but no Ba ($Z=56$), attributed to an aspherical hypernova \citep{Ezzeddine2019}.
SMSS J2003$-$1142 has a full r-process pattern from Sr to Eu (and beyond), attributed to a magnetorotationally driven hypernova \citep{Yong2021}.
AS0039 is a star in the Sculptor dwarf galaxy also suggested to be consistent with a high energy but spherical hypernova, and it has very low Sr and Ba \citep{Skuladottir2021}.
It appears that hypernovae are able to generate a whole range of neutron-capture nucleosynthesis, as might be expected based on potentially variable strengths of the central engine.
Our star \TheStar has a full complement of first neutron-capture peak elements from Sr to Pd and nothing beyond, which makes it most similar to HE1327$-$2326's much sparser abundance pattern. Note that the pattern is flat in [X/Fe], which differs substantially from the ``pure'' r-process pattern in SMSS J2003$-$1142 as well as theoretical r-process predictions \citep{Holmbeck2023}.

\subsection{Comparison to Typical Metal-Poor Stars}

\begin{figure}
    \centering
    \includegraphics[width=0.8\linewidth]{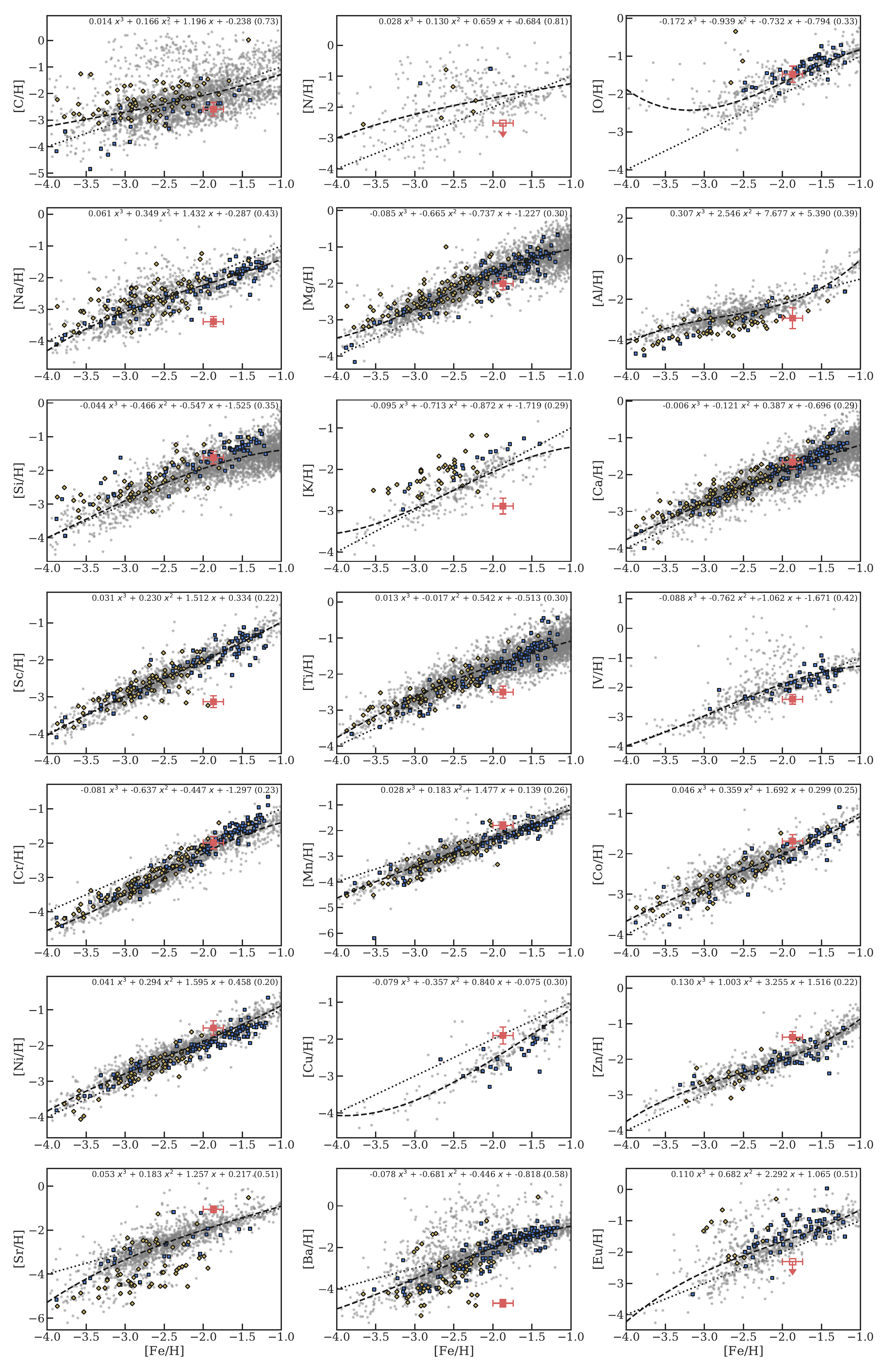}
    \caption{[X/H] vs [Fe/H] for SAGA (grey points) and dwarf galaxy stars (blue is classical dSph, yellow is ultra-faint dwarf). 
    \TheStar is shown as a large red square.
    The dotted line indicates [X/Fe]=0, and the dashed line is the best-fit to the SAGA stars that we use as an empirical ISM composition. Note that the comparison abundances have been shifted by the NLTE and evolutionary corrections for \TheStar.}
    \label{fig:saga_ism_xh_all}
\end{figure}

As a complement to Figures~\ref{fig:xfe_saga} and \ref{fig:fehxh}, Figure~\ref{fig:saga_ism_xh_all} shows [Fe/H] vs [X/H] for 21 elements. The grey points are halo stars from the SAGA database \citep{Suda2008}, and blue and yellow points are an extended dwarf galaxy compilation from JINAbase \citep{jinabase}. The literature abundances have been shifted by the NLTE and evolutionary corrections from Table~\ref{tab:abunds}.
We show the one-to-one line ([X/Fe]=0) as a dotted black line, and an outlier-clipped polynomial fit to the SAGA abundances as a dashed black line (best-fit indicated on each panel), which represents the typical ISM composition at any given metallicity. 
At its [Fe/H], \TheStar is a visible low outlier in the Na, Al, K, Sc, Ti, V, Ba, and Eu panels, while simultaneously being a high outlier in Mn, Co, Ni, Cu, Zn, and Sr.

\section{Nucleosynthesis Origin}\label{app:nuc}

We performed an extensive literature search for nucleosynthesis predictions that could match \TheStar. In this section, we will mostly ignore the metallicity constraint from \TheStar (Section~\ref{sec:dilution}) and instead focus on what sites or conditions could produce the observed abundance pattern.

\subsection{Brief nucleosynthesis summary}\label{app:nucsummary}

We start by briefly summarizing some key element ratios and the main physics of the supernova progenitor and explosion that drives their values, discussing them in the context of \TheStar. 
\begin{itemize}
    \item The [C/O] value is a good indicator of the zero-age main-sequence mass of the progenitor, where a lower value indicates a more massive star \citep[e.g.,][]{Ishigaki2018}.
    The low [C/O] value of \TheStar is typically found for yields of massive progenitors with ZAMS mass $\gtrsim$20$M_{\odot}$ \citep{Heger10,Nomoto2013}.
    We note that this ratio and the overall products of carbon burning are subject to uncertainties in the $^{12}$C($\alpha$,$\gamma$)$^{16}$O rate and the treatment of convection \citep[e.g.,][]{Imbriani2001,ElEid2004,deBoer2017,Farmer2019}.
    Additionally, binary interactions may affect [C/O] predictions, though current models in solar-metallicity stars suggest interactions increase C yields \citep{Farmer2021,Farmer2023ApJ}.
    
    \item{} [N/O] can potentially help constrain rotation in the progenitor, as nitrogen is typically enhanced in rotating stars \citep[e.g.,][]{Choplin2018AA}. However, while \TheStar has a very low upper limit on the abundance of nitrogen, we found it is insufficient to rule out rotating metal-poor progenitors, as the large N enhancement would still be below our detection threshold \citep{Ekstrom2008AA,Limongi2018}.
    
    \item The extreme odd-even effect seen for elements from C to Sc, characterized by low values of [Na/Mg] and [K/Ca], occurs in stars with low neutron-fractions. Such values are typically associated with PISNe \citep{Heger2002ApJ, Kozyreva2014AA, Takahashi2018} but also occur to a lesser extent in any massive zero or low-metallicity progenitor \citep{Heger10,Nomoto2013,Limongi2018}.
    The low [Na/Mg] especially prefers higher initial mass progenitors with $M > 70 \msun$.
    Rotation can ``fill in'' the odd elements as well \citep[e.g.,][]{Choplin2018AA}.
    
    \item The low alpha abundances e.g. [Mg/Fe] indicate a low ratio of hydrostatic to explosively synthesized elements, which tends to occur either in lower mass CCSN progenitors ($\lesssim 15 \msun$, \citealt{McWilliam13,Carlin18}) or also in the most massive PISNe \citep{Heger2002ApJ,Salvadori2019}.

    \item The composition of the iron group from Sc to Zn depends sensitively on the details of the supernova explosion, such as explosion energy, remnant mass, convection, jets, and more.
    All of these elements are produced during explosive silicon burning, with different degrees of contribution from complete and incomplete burning and likely important 3D effects \citep[e.g.,][]{Curtis19,Sieverding2023}.
    \TheStar has an unusual iron group composition, showing very low [Sc, Ti, V/Fe] along with high [Mn, Ni, Zn/Fe].
    We were unable to find any existing supernova yield model matching the whole Fe peak pattern.
    
    \item The elements from Sr and heavier are formed primarily through the slow (s-) and rapid (r-) neutron-capture processes, though proton-capture and i-processes are possible as well.
    In \TheStar, the very low [Ba/Fe] rules out a strong, neutron-rich r-process \citep[e.g., from neutron star mergers,][]{Holmbeck2023} and the main s-process \citep[e.g.,][]{Lugaro12}.
    Most remaining scenarios to explain the high [Sr--Pd/Fe] invoke nucleosynthesis associated with the formation of neutron stars or black holes, with the i-process activated in external He-rich layers as an alternative.
    
\end{itemize}

\begin{figure}
    \centering
    \includegraphics[width=0.95\linewidth]{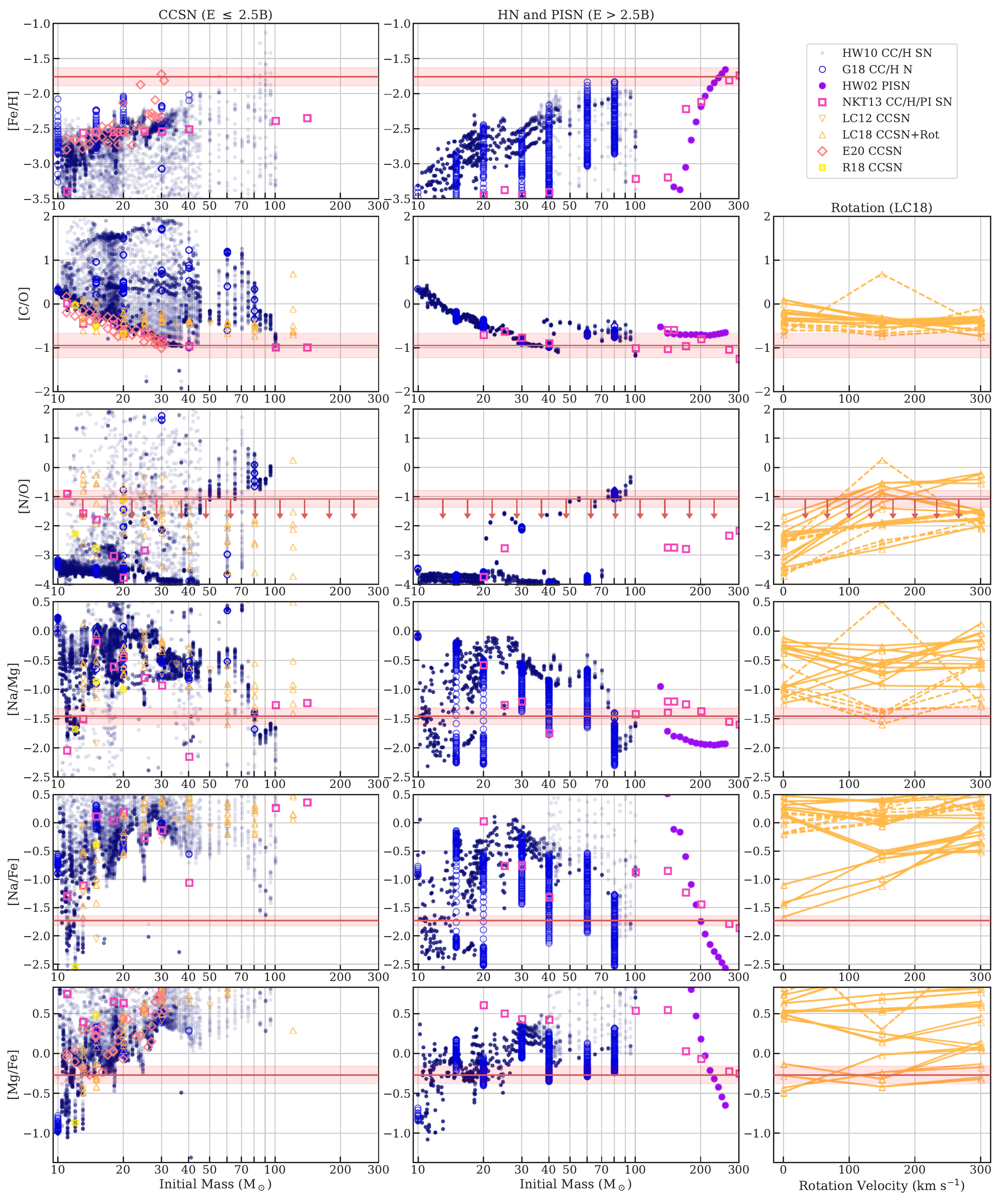}
    \caption{Light element ratios vs initial mass for CCSN (left column; HW10, G18, NKT13, LC12, LC18, E20, R18), HN and PISN (center column; HW10, G18, NKT13, HW02), and rotation velocity (right column, LC18 only).
    N and Na are not plotted for E20.
    In the right column, the same mass and metallicity at different rotation velocities are connected by solid lines ($<50\msun$) or dashed lines ($\geq 50\msun$).
    These figures explain the first rows of Table~\ref{tab:death}.}
    \label{fig:deathgrid_1}
\end{figure}

\begin{figure}
    \centering
    \includegraphics[width=0.95\linewidth]{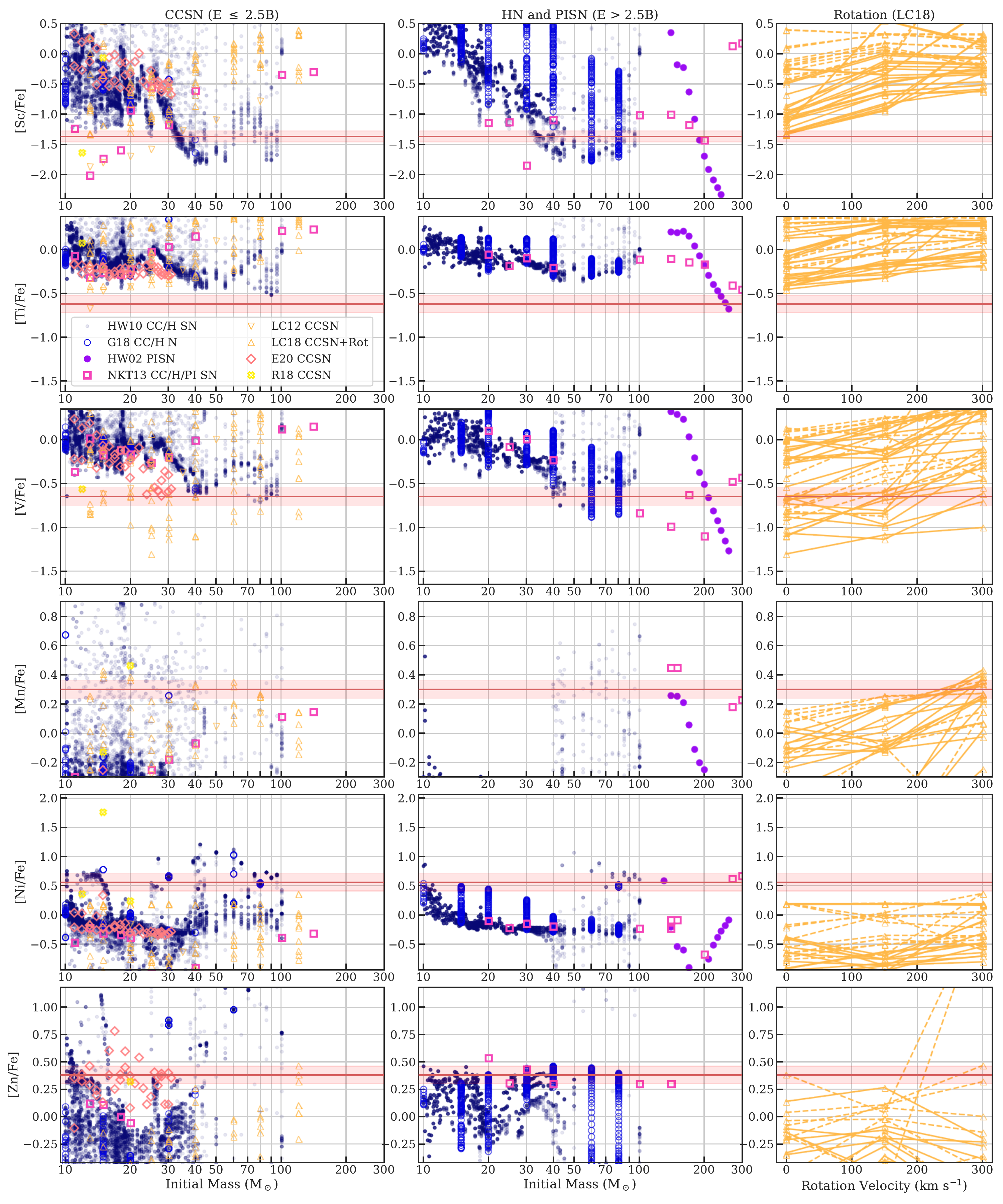}
    \caption{Iron peak element ratios vs initial mass for CCSN (left column; HW10, G18, NKT13, LC12, LC18, E20, R18), HN and PISN (center column; HW10, G18, NKT13, HW02), and rotation velocity (right column, LC18 only).
    In the right column, the same mass and metallicity at different rotation velocities are connected by solid lines ($<50\msun$) or dashed lines ($\geq 50\msun$).
    These figures explain the middle rows of Table~\ref{tab:death}.}
    \label{fig:deathgrid_2}
\end{figure}

In Figures~\ref{fig:deathgrid_1} (light elements) and \ref{fig:deathgrid_2} (Fe peak elements), we show plots of all the key element ratios with respect to progenitor initial mass and explosion energy for eight models that span the range of predictions: primordial supernovae from \citet{Heger2002ApJ} (HW02), \citet{Heger10} (HW10), \citet{Limongi2012} (LC12), \citet{Nomoto2013} (N13), \citet{Grimmett2018MNRAS} (G18), and \citet{Ebinger2020ApJ} (E20); and low metallicity supernovae from \citet{Limongi2018} (LC18, including rotation and keeping the most massive stars that may not explode), \citet{Ritter2018} (R18), as well as NKT13, E20.
We somewhat arbitrarily put a line at 2.5B to split models between CCSN and HN.
We cut models to only those with metallicity $<-1.5$ solar.
Note that the models by LC12, LC18, and R18 do not have explosion energies provided, so we do not include those on the explosion energy figures; and the progenitor models in E20 did not include odd elements, so we removed it from the N and Na panels for Fig~\ref{fig:deathgrid_1}.
The horizontal red lines and shaded region indicate the observed value and uncertainty in \TheStar, including an upper limit for [N/O].

The comparison between these figures and \TheStar are qualitatively summarized by Table~\ref{tab:death}.
Green check marks are given when nearly all models in a particular category can match \TheStar.
We add use a check mark with a question mark if existing models would work, but we felt there were large theoretical uncertainties or important exceptions.
Red X marks indicate that current theoretical models suggest a particular criterion is impossible for some category of models.
The orange X with question mark indicates that most models do not satisfy the criterion, but we felt there were large theoretical uncertainties and/or large variations in existing predictions.
The ``rapid rotation'' column uses the LC18 conclusions, though the behavior could potentially change in other models.
Overall, we can see that the $50-100\msun$ HN column has no observation fully ruling it out, although there are substantial uncertainties.
The massive PISN column has the most solid check marks, but it is solidly ruled out by the Fe peak and neutron-capture elements.
(The neutron-capture elements will be discussed more in Appendix~\ref{app:nucncap}.)

A few rows in Table~\ref{tab:death} merit more discussion.
We originally expected the low [N/O] limit to substantially constrain rotation velocities or initial masses \citep{Meynet06,Ekstrom2008AA,Placco16}, but it turns out that since [N/O] is so low in most metal-free SNe, even a 100x increase is not that constraining.
However increased rotation does substantially impact [Na/Fe], as well as all the odd Fe peak elements. It is also strongly constrained by the low Ba, since the s-process in fast-rotating massive stars would increase Ba substantially \citep{Chiappini2013,Choplin2018AA}.

We see that Fe peak synthesis is very uncertain in both CCSNe and HNe, reflecting uncertainties in the explosion mechanism.
The lighter Fe peak elements (Sc, V) tend to prefer higher mass CCSNe or HNe, and almost no CCSN or HN models can explain the low [Ti/Fe].
The high Mn is problematic: for CCSN and HN, it can only be matched by lower mass, higher metallicity progenitors with rapid rotation or fine-tuned mixing/fallback (which would not produce high [Fe/H] ratios). However those same LC18 models also greatly overpredict the Sc, Ti, and V abundances. The low-mass PISNe also produce the appropriate [Mn/Fe], but only because they produce so little Fe.
Higher explosion energies ($E > 10$B) can produce high [Zn/Fe] ratios, but Ni is difficult overall except in lower mass CCSNe.

\begin{table*}
    \caption{Death Matrix\label{tab:death}}
    \hspace*{-2.8cm}
    \begin{tabular}{c|ccc|ccc|cc|c}
    \tableline \tableline
    Criterion  & \multicolumn{3}{c|}{CCSN} & \multicolumn{3}{c|}{HN} & \multicolumn{2}{c|}{PISN} & Add Rapid \\
    Mass (\msun) & $<$20 & 20--50 & 50--100 & $<$20 & 20--50 & 50--100 & 140--200 & 200--260 & Rotation \\
    \tableline
    High [Fe/H] $\sim -1.8$ & \cxmark  & \cxmarkq & \cxmarkq & \cxmark  & \cxmark  & \cxmarkq & \cxmark  & \ccmark & \nodata \\ \tableline
    Low [C/O]    $\sim-1.0$ & \cxmark  & \ccmarkq & \ccmarkq & \cxmark  & \ccmark  & \ccmark  & \ccmark  & \ccmark & No change \\
    Low [N/O]      $< -1.0$ & \ccmarkq & \ccmarkq & \cxmarkq & \ccmark  & \ccmark  & \ccmarkq & \ccmark  & \ccmark & Increase up to 100x \\
    Low [Na/Mg]  $\sim-1.5$ & \ccmarkq & \cxmark  & \ccmarkq & \ccmarkq & \ccmarkq & \ccmark  & \ccmark  & \ccmarkq& Varying predictions \\
    Low [Mg/Fe]  $\sim -0.3$& \ccmark  & \cxmark  & \cxmark  & \ccmark  & \ccmarkq & \ccmarkq & \cxmark  & \ccmark & Little change \\
    \hline 
    Low [Sc,Ti,V/Fe]  $< -0.5$ & \cxmarkq & \ccmarkq & \ccmarkq & \cxmark  & \ccmarkq & \ccmarkq & \cxmark  & \ccmark & Increase by 2-10x \\
    High [Mn/Fe]    $\sim 0.3$ & \cxmarkq & \cxmarkq & \cxmarkq & \cxmarkq & \cxmarkq & \cxmarkq & \ccmarkq & \cxmark & Increase by 2-3x \\
    High [Ni,Zn/Fe] $\sim 0.5$ & \ccmarkq & \cxmarkq & \cxmarkq & \ccmarkq & \ccmarkq & \ccmarkq & \cxmark  & \cxmark & Little change \\
    \hline 
    High [Sr--Pd/Fe] $> 0.5$ & \ccmarkq & \ccmarkq & \ccmarkq & \ccmarkq & \ccmarkq & \ccmarkq & \cxmark  & \cxmark & Increases all \\
    Low [Ba/Fe]  $\sim -2.8$ & \ccmark  & \ccmark  & \ccmark  & \ccmark  & \ccmark  & \ccmark  & \ccmark  & \ccmark & Increases Ba \\
    \hline 
    Explosion Expected & \ccmark  & \ccmarkq & \cxmark  & \ccmark  & \ccmarkq & \ccmarkq & \ccmark  & \ccmark & Helps explodability. \\
    \tableline
    \end{tabular}
    \tablecomments{
    The \ccmark$~$ and \cxmark$~$ signify whether nearly all models in a particular category are found to match or fail to match a given criterion.
    The \ccmarkq denotes that most models in a category satisfy the criterion, but there are large uncertainties and/or important exceptions.
    Similarly, the \cxmarkq denotes that most models do not satisfy the criterion, but there are large uncertainties and/or large variations in predictions.
    All Fe peak predictions for CCSNe and HNe have significant uncertainties.
    The last row indicates whether an explosion is theoretically expected in this mass range for metal-poor progenitors.
    }
\end{table*}
\subsection{Search Through Single Star Nucleosynthesis Yields}\label{app:nucsingle}

\input{nucref}

We performed an extensive search of supernova yield grids covering a wide range of possible nucleosynthesis sites, covering elements from C to Zn:
primordial core-collapse supernovae exploded with a piston \citep[][assuming the S4 location]{Heger10}, thermal bomb \citep{Nomoto2013}, and kinetic bomb \citep{Limongi2012} with various assumptions for mixing and fallback;
core-collapse supernovae of higher metallicities \citep{Nomoto2013,Ritter2018}, with rotation \citep{Limongi2018} and engine-driven explosions \citep{Ebinger2020ApJ};
hypernovae of varying energies \citep{Nomoto2013,Grimmett2018MNRAS};
and primordial pair instability supernovae \citep{Heger2002ApJ,Nomoto2013}.
We also examined core-collapse supernovae of solar metallicity binary stripped stars \citep{Farmer2023ApJ} and thermonuclear Type Ia supernovae \citep[references in][]{Reggiani2023}, though none of these were good fits so we do not discuss them further.
We only included models with [Z/H] $< -1.5$, except the binary stripped star supernovae where only solar metallicity models exist.
We also qualitatively considered how abundance patterns would be affected by nucleosynthesis of jets interacting with stellar envelopes \citep{Grimmett2021MNRAS} and rotation \citep{Ekstrom2008AA,Frischknecht16}, as these references did not provide full yield tables of all elements.
A summary of the searched models is given in Table~\ref{tab:nucref}.

\begin{figure}
    \centering
    \includegraphics[width=0.95\linewidth]{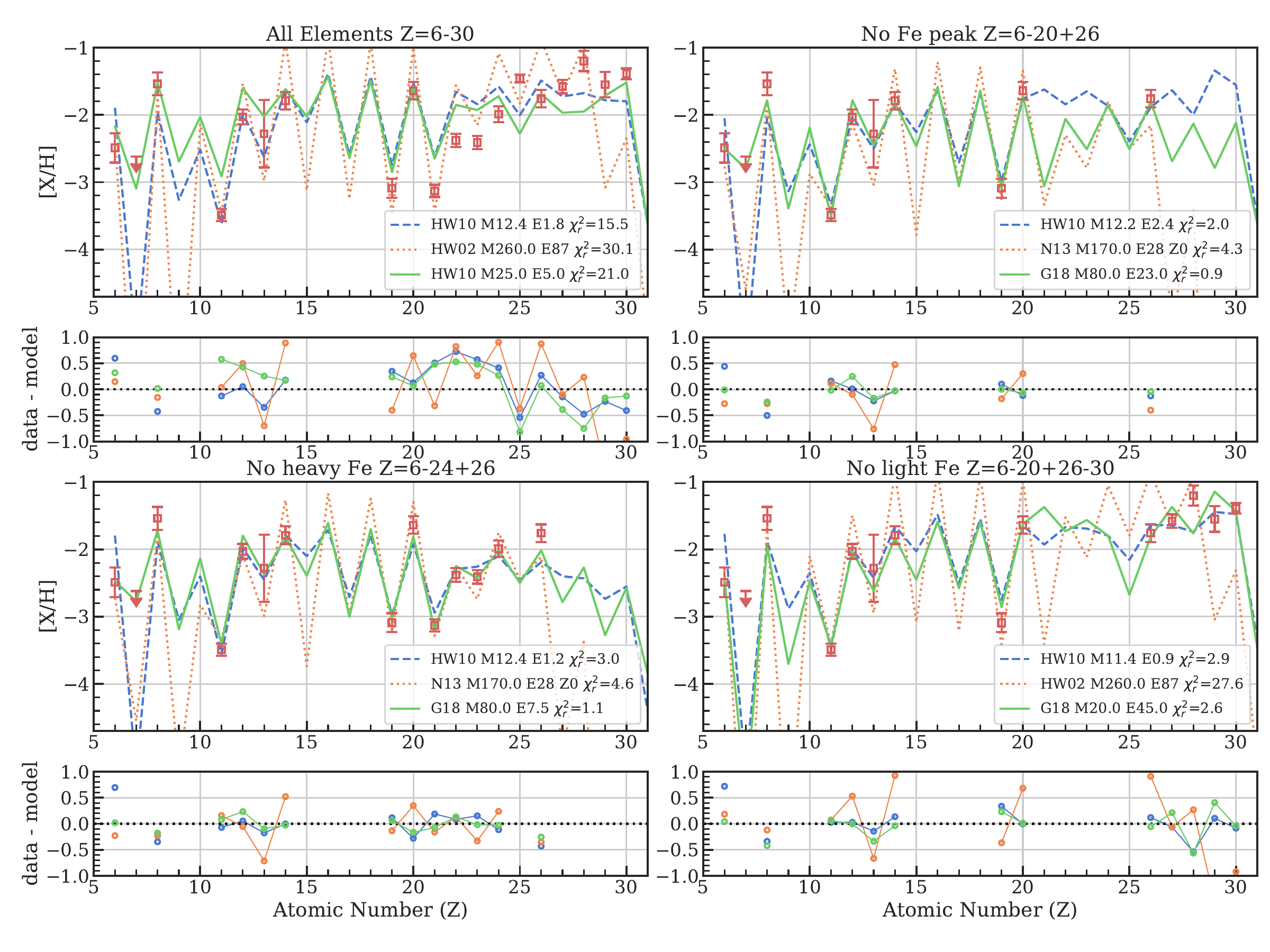}
    \caption{
    Results of grid search through CCSN, PISN, and HN models for different permutations of elements.
    The best fit model is shown in the legend,
    dashed blue = CCSN, dotted orange = PISN, solid green = HN.
    Reduced $\chi^2$ is shown in the legend. See text for details.
    }
    \label{fig:singlestarfit}
\end{figure}

To investigate the yield grids, we used two search algorithms: \code{Starfit} which does $\chi^2$ minimization, \citealt{Heger10}; and a code based on \citet{Ji2020a} that minimizes the mean absolute deviation (normalized by abundance uncertainty), which is more robust to outliers. In both searches, the gas dilution is a free parameter, i.e. we allow the abundance pattern of each model to shift arbitrarily up and down in log space, ignoring the total metallicity constraint that would rule out essentially all models. The two different search strategies resulted in the same overall conclusions.

No model or site was found to adequately reproduce the entire observed abundance pattern, so the best fit models depended heavily on the choice of elements to fit.
After examining many permutations, it became clear that the lighter elements with $Z \leq 20$ could generally be fit by many types of SNe, but the iron peak was never well-fit.
We thus present the best-fit CCSN, HN, and PISN models for four permutations of elements in Figure~\ref{fig:singlestarfit}:
(1) including all elements from $Z=6-30$,
(2) fitting elements from $Z=6-24$ and $Z=26$ to exclude the enhancements in heavier Fe peak elements as well as the difficult element Mn,
(3) fitting elements from $Z=6-20$ and $26-30$ to remove the deficiency of lighter Fe peak elements, and
(4) fitting $Z=6-20$ and $Z=26$ to remove all Fe peak elements other than iron.

When fitting all elements from C to Zn (top left panel), the best fits were achieved by low mass SN,  12.4\msun primordial SN from \citet{Heger10}, but with a reduced $\chi^2$ of 15.5 that indicates a terrible fit. This model is able to explain elements from Na-Si and Ca, but it fails C and O and the entire Fe peak. Other CCSN models have similar issues.
If we remove all Fe peak elements other than Fe (top-right panel), we see that there is no CCSN or PISN able to reproduce all the light elements. Low-mass CCSN fail due to C/O as shown, while high-mass ordinary energy CCSN produce too little Fe. The moderate-sized odd-even effect in \TheStar suggests lower mass PISNe, but these produce insufficient Fe. However, a massive (80 \msun) high-energy (23B) hypernova from \citet{Grimmett2018MNRAS} is able get a near-perfect fit to the light elements while producing enough iron ($\chi^2_r=0.9$).
This scenario does not change much if we add back in Sc, Ti, V, and Cr (bottom left panel), where the CCSN and PISN continue to have issues but the 80 \msun HN still works well ($\chi^2_r=1.1$).
However if instead we exclude the light Fe peak and add back the heavy Fe peak (bottom right), again no model gives a satisfactory fit.

This exploration suggests that if a single star is to produce the abundance pattern, the best candidate is an 80 \msun hypernova. This can explain all the light elements from C to Cr, and the exact Fe peak can be adjusted based on the energy.
Plausibly, fixes to the heavy Fe peak pattern and enhanced Mn could come from nucleosynthesis in jets and/or induced by rotation or other 3D effects, which are not extensively explored in existing model grids.
However, current nucleosynthesis models suggest this does \emph{not} work, as jets tend to co-produce Sc/Ti/V and Co-Zn \citep{Tominaga09,Grimmett2021MNRAS}.
Additionally, all these higher energy HN models violate the metallicity constraint (Section~\ref{sec:dilution}).
One other option is adding i-process to PISNe, which could resolve some of the problems with those models (see Appendix~\ref{app:nucncap}).

\subsection{Combining Two Sites}\label{app:nucbinary}

\begin{figure}
    \centering
    \includegraphics[width=\linewidth]{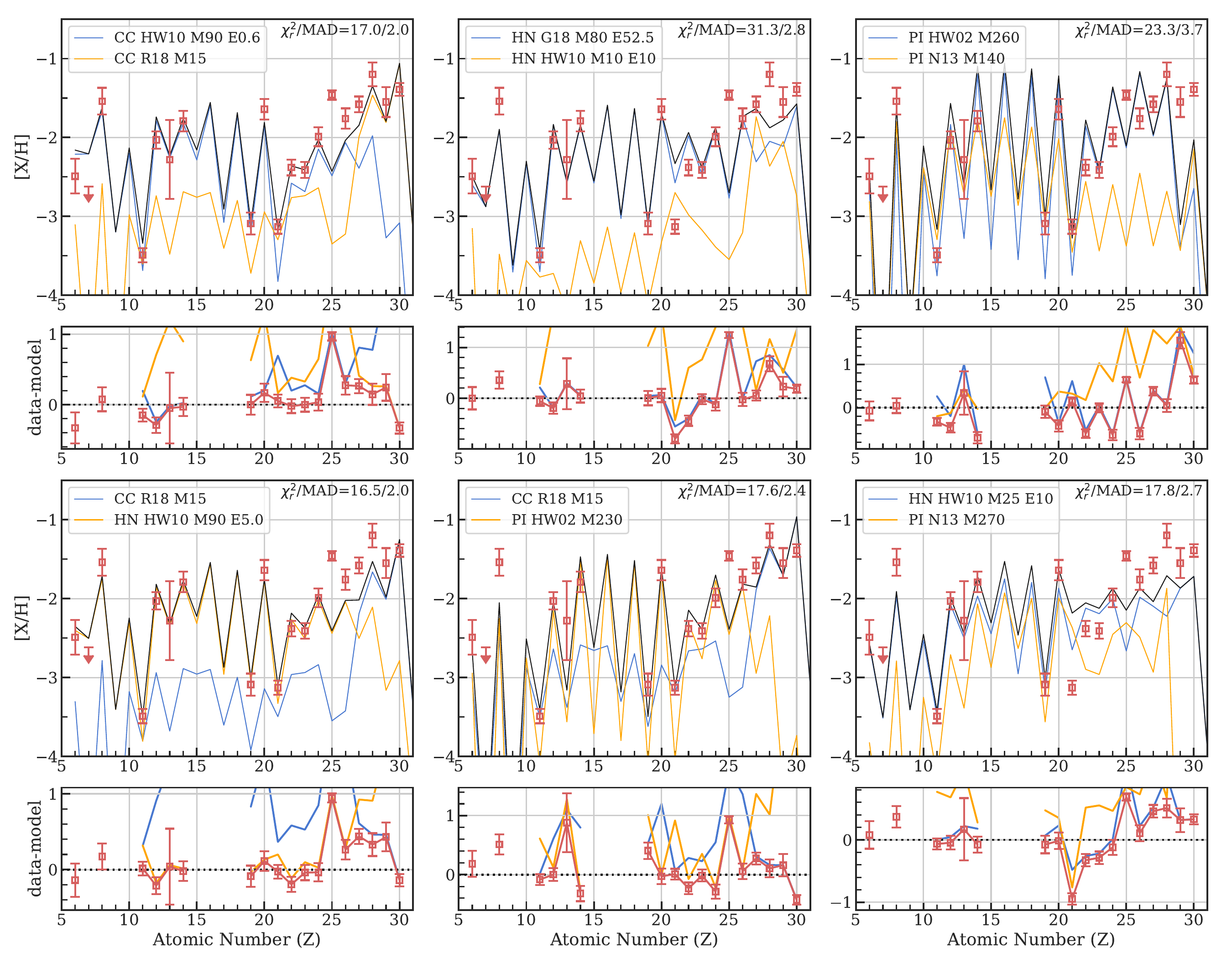}
    \caption{Best-fit results of fitting abundances of \TheStar (red points) combining CCSN, HN, and PISN with each other. The large panels shows the fits in [X/H], while the smaller panels indicates the residual (data - model).
    The blue and orange lines indicate two different models described in the upper left legend of each panel, and the black line indicates their sum. The top-right of the large panels shows reduced chi squared and the error-normalized mean absolute deviation.
    Reference Key: E20 \citep{Ebinger2020ApJ}, G18 \citep{Grimmett2018MNRAS}, HW02 \citep{Heger2002ApJ}, HW10 \citep{Heger10}, N13 \citep{Nomoto2013}, R18 \citep{Ritter2018}.
    See Section~\ref{app:nucbinary} for discussion.
    }
    \label{fig:best_binary_1}
\end{figure}

\begin{figure}
    \centering
    \includegraphics[width=\linewidth]{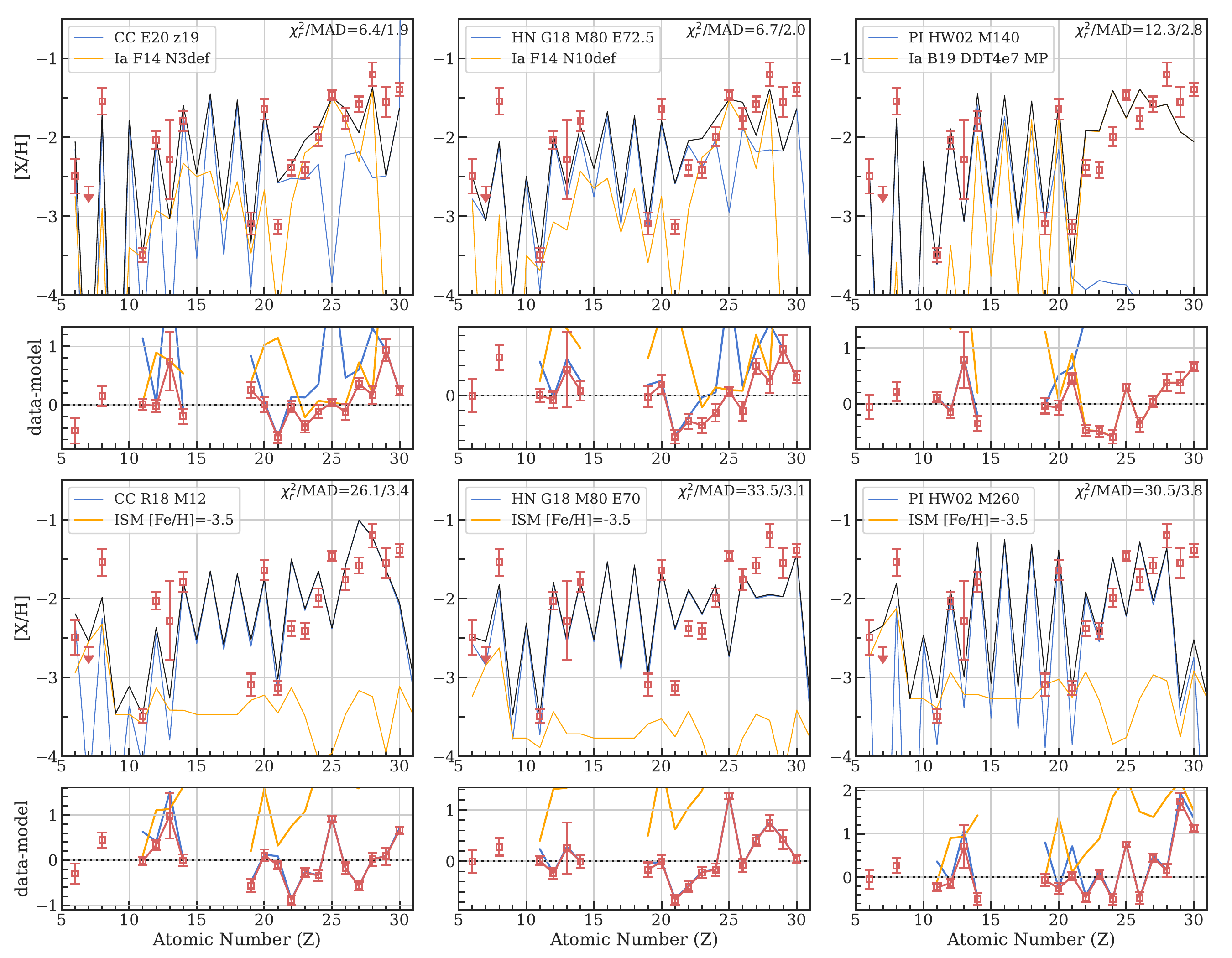}
    \caption{Best-fit results of fitting abundances of \TheStar (red points) combining CCSN, HN, and PISN with SN Ia and ISM. Same lines as Fig~\ref{fig:best_binary_1}.
    Reference Key: B19 \citep{Bravo2019MNRAS}, E20 \citep{Ebinger2020ApJ}, F14 \citep{Fink2014}, G18 \citep{Grimmett2018MNRAS}, HW02 \citep{Heger2002ApJ}, R18 \citep{Ritter2018}.
    See Section~\ref{app:nucbinary} for discussion.
    }
    \label{fig:best_binary_2}
\end{figure}

As no individual site provided a satisfactory fit to the entire abundance pattern from C to Zn ($Z=6-30$), we next ran joint fits between all possible pairs of sites.
Note that combining two sites is not a plausible explanation for \TheStar (see Section~\ref{sec:dilution}), but this exploration can identify physical conditions in existing calculations that could explain parts of the abundance pattern if combined into one site.
It is expensive to consider all pairs and there is no analytic solution allowing arbitrary dilution, so we approximated the solution by first fitting each yield to the star individually following \citet{Ji2020a}, then performing a brute force optimization of the minimum absolute deviation (normalized by abundance error) by allowing the dilution of each yield to drift by $-3.0$ to $+1.0$ dex in units of 0.1 dex from the initial fit.
For simplicity we ignored the N upper limit.

For this exercise, we split our yield tables into CCSN (CC), HN, PISN (PI), and SN Ia (IA). The \citet{Heger10} and \citet{Grimmett2018MNRAS} yield tables spanned a large range of energies, so we somewhat arbitrarily split them into CC and HN at $E=2.5$B.
We also created an empirical ISM model using abundances from SAGA \citep{Suda2008}, fitting third order polynomials to [Fe/H] vs [X/H] as shown in Figure~\ref{fig:saga_ism_xh_all} and assuming [X/Fe]$=0$ for elements not in that figure. We created ISM compositions at $\feh=-3.5$ to $-1.5$ in steps of 0.5 dex and included them as a separate yield table in our fits.

The results are shown in Figures~\ref{fig:best_binary_1} and \ref{fig:best_binary_2}.
Fig~\ref{fig:best_binary_1} shows a search for the optimal combination of CC$+$CC, HN$+$HN, and PI$+$PI (top row) and 
crossing SN types CC$+$HN, CC$+$PI, and HN$+$PI (bottom row).
Fig~\ref{fig:best_binary_2} crosses each SN type with SN Ia (top row) and ISM (bottom row).
These specific fits were chosen to have the smallest possible minimum absolute deviation of all combinations. Close ties were broken by choosing models that illustrate specific physics.

The immediate overall conclusion is that no pairs of models can fit the entire abundance pattern, despite the vastly expanded model space compared to single site fits. The reduced $\chi^2$ and mean absolute deviation (MAD) should be near 1 for reasonable fits, and the lowest MAD is 1.9.
There is enough freedom that the light elements with $Z \leq 20$ are generally fit well. However, the Fe peak from $Z=21-30$ has many issues, especially for Sc ($Z=21$), Mn ($Z=25$), and the overall enhanced Co-Zn ($Z=27-30$). Fitting all three of these simultaneously is not possible even after searching through over 20,000 yield sets. 

Two models show up repeatedly in these fits as being able to solve some of the Fe peak problems.
First, in the top row of Fig~\ref{fig:best_binary_2}, off-center deflagration explosions of Chandrasekhar mass white dwarfs (here showing yields from \citealt{Fink2014}) are able to achieve a high [Mn/Fe] and [Ni/Fe] ratio that broadly matches \TheStar, though no SN Ia is able to produce similarly high levels of Zn. Though white dwarfs are likely not relevant progenitors for nucleosynthesis in \TheStar (see Section~\ref{sec:dilution}), the high neutron fractions achieved in those white dwarfs could potentially be achieved in massive stars, although there is tension with the low neutron fractions needed to have a large odd-even effect.
Second, the 15 \msun CC model from \citet{Ritter2018} shows up several times in both figures because it is able to produce large amounts of Co-Zn. This model ejects a large amount of material near the mass cut that was in nuclear statistical equilibrium (NSE), resulting in a very strong alpha-rich freezeout that produces this pattern, as well as some heavier elements all the way up to Mo ($Z=42$; see fig. 28 of \citealt{Ritter2018}, \citealt{Woosley1992}). Ejecting substantial amounts of this NSE material is a promising future path to reproducing the abundance pattern of \TheStar. One example of this could be late-time mass loss induced by close binary interactions. The lighter elements (C-Ca) could be hydrostatically synthesized by a $80 \msun$ progenitor to match \TheStar, but late mass loss from binary stripping after the CO core mass is set could push the mass cut closer to the center of the star. Note that Case C mass transfer (after igniting central He burning) is a typical outcome of massive metal-poor star binary evolution \citep{deMink2008}.

Finally, the models including ISM did not work well. As expected, the ISM in the PISN+ISM was able to fix the overly strong odd-even effect in the light elements for the 260\msun PISN, but it could not contribute to the problems with the Fe peak. Though we allowed the ISM to be metallicities from $\feh=-3.5$ to $-1.5$, the $-3.5$ ISM always came to the forefront primarily because it helped most models fit the [C/O] ratio better while also filling in any missing amount of Na, K, or Sc. It is worth noting that the [C/O] ratio in our ISM model is likely too low, since it is only possible to detect O in the data when it is highly enhanced (and similar for N).

\subsection{Full fits from $Z=6-56$ including neutron capture elements}\label{app:nucncap}

There are few models that self-consistently predict nucleosynthesis yields of both lighter and heavier elements.
We thus constructed plausible combinations of our best-fit explosion models (HN and PISN) and heavy-element nucleosynthesis patterns (computed through simulation-based or parameterized trajectories), combining them using the two-component search from \code{Starfit} \citep{Heger10}. Our goal was to create realistic abundance patterns for scenarios that could simultaneously explain both the light and heavy elements in \TheStar.
These results were used to create the models including neutron-capture elements in Fig~\ref{fig:nucl_best}.
A PISN combined with an $i$-process emerges as one possible explanation for simultaneously producing the observed abundance pattern and high metallicity of \TheStar, although no such theoretical models currently exist in the literature. A hypernova combined with an $r/i$-process or a neutrino-driven wind (if such an explosion occurs) may also provide a good match to the observed abundances. In this section we elaborate on general considerations for heavy element nucleosynthesis, followed by discussion of which combinations are realistic.

Considering the heavy elements in \TheStar ($Z > 35$), the enhancement in Sr all the way out to Pd, combined with low Ba and Eu, provides a strong constraint on the conditions needed for their synthesis. There will be insignificant contributions from the main r-process that synthesizes a relatively large amount of Ba and Eu \citep[e.g.][]{Holmbeck2023} and the main s-process that synthesizes a large amount of Ba relative to Sr in metal-poor stars \citep[e.g.,][]{Lugaro12}. The i-process generally has a similar challenge as the s-process of overproducing Ba \citep[e.g.,][]{Hampel2016,Cote2018}. This is mostly driven by the assumption of solar scaled initial abundances for stellar simulations at low metallicity, where an early increase of the Ba production is mostly fed from seed abundances in the Sr-Zr region. The i-process production at the first neutron-magic peak of Rb-Zr is instead due to neutron captures on the Fe seeds \citep[e.g.,][]{Herwig2011}, and it does not depend on their initial abundances.
Finally, most $\nu$-driven wind models only produce significant amounts of elements up to Mo but are unable to extend out to Pd \citep{Frohlich2006PRL,Wanajo2006,Wanajo2018ApJ}.

From published model grids, we find that a weak r-process with $Y_e\sim$0.25 and entropy of 12$k_B$/baryon (or something slightly more neutron-poor) allows the synthesis of the first-peak elements without producing a substantial amount of Ba \citep[e.g.,][]{Nishimura2017ApJ,Holmbeck2023}. The detailed pattern of any individual $Y_e$ trajectory does not match our observations, but a mixture of ejecta conditions is expected and can likely be combined to reproduce the detailed pattern \citep[e.g.,][]{Farouqi10,Holmbeck2019}.
We also find that a high entropy (120 $k_B$/nucleon) and proton-rich ($Y_e = 0.54$) neutrino-driven wind trajectory is able to match the light elements out to Pd, although it is not clear if such conditions can be achieved in actual supernovae \citep{Bliss2018ApJ}.

In Section~\ref{sec:discussion} we discussed a computational experiment for i-process nucleosynthesis where we reduced abundances of elements heavier than Fe. This is justified as metal-poor gas is deficient in the neutron-capture elements compared to iron (see Figure~\ref{fig:saga_ism_xh_all} and \citealt{Cescutti13}). Our i-process models thus use the nucleosynthesis framework from \citet{Bertolli2013} and \citet{Roederer2016c,Roederer2022}, but reduce the initial heavy element abundances with respect to Fe by ten times (e.g., $\mbox{[Sr/Fe]}=-1$ and $\mbox{[Ba/Fe]}=-1$). With this change, it is possible to satisfy the low Ba constraint and at the same time obtain an efficient production in the Rb-Ru mass region as observed in \TheStar. However, while the i-process produces heavy elements, the abundances of intermediate mass elements (in particular the less abundant odd elements) can also be affected. An example is the observed enhancement of Sc with respect to Ca in the post-AGB star Sakurai's object, where Sc is co-produced with heavy elements in the Rb-Zr region \citep{Herwig2011}. Another example is extra production of Na and Al \citep[e.g.,][]{Clarkson18}.

We now discuss whether the production of heavy elements through the channels described above is a realistic possibility in CCSN, HN, and PISN explosions. CCSNe/HNe are thought to be accompanied by heavy element production in neutrino-driven winds from a NS remnant, either through a weak r-process or under proton-rich conditions, although the detailed properties of these winds are being investigated \citep{Pruet2006ApJ, Frohlich2006PRL, Arcones11, Arcones2013JPhG, Wanajo13, Fujibayashi2015ApJ}. A weak r-process with $Y_e\lesssim$0.25 could potentially occur in accretion disk winds around a black hole \citep[e.g.,][]{Pruet2004,Surman2006}. Altogether, the production of enhanced light neutron-capture elements out to Pd in such winds suggests scenarios involving material ejected from a neutron star or black hole accretion disk. These remnants are a natural and expected outcome of CCSN/HN explosions, lending support to the scenarios where a HN is combined with a weak r-process or proton-rich neutrino-driven wind. On the other hand, PISNe leave no remnant behind, and hence cannot produce neutron-capture elements through these channels. 

The other possibility for producing heavy elements, the intermediate (i-) neutron-capture process, is a recently revived process of interest \citep[e.g.,][]{Hampel2016,Roederer2016c,Cote2018}.
The source of i-process  in massive metal-poor stars is proton ingestion due to mixing with convective He shells \citep{Cowan1977,Banerjee18a,Clarkson18}. However, it is not yet clear if 3D simulations including convection are able to trigger the proton ingestion that would induce the i-process \citep{Herwig2014,Woodward2015}. Nonetheless, this possibility lends support to the scenario where a HN combined with an i-process could explain the abundances in \TheStar.

If massive metal-poor progenitors of CCSNe or HNe can be affected by proton ingestion events that activate the i-process, we can expect that the same may happen for PISN progenitors. If this occurs, the i-process in a PISN progenitor could not only produce the neutron-capture elements, but it might also help fill in the overly low light odd element abundances, while also increasing the abundance of Zn that otherwise rules out PISN models. However, there are no PISN models available yet taking into account the impact of such events within their integrated yields. Future generations of PISN models including proton ingestion and the supernova explosion will be paramount to explore this scenario in greater detail.




\end{document}

%% file: authors.tex
\author[0000-0002-4863-8842]{Alexander~P.~Ji}
\affiliation{Department of Astronomy \& Astrophysics, University of Chicago, 5640 S Ellis Avenue, Chicago, IL 60637, USA}
\affiliation{Kavli Institute for Cosmological Physics, University of Chicago, Chicago, IL 60637, USA}
\affiliation{Joint Institute for Nuclear Astrophysics -- Center for Evolution of the Elements (JINA), East Lansing, MI 48824, USA}
\author[0000-0002-3211-303X]{Sanjana~Curtis}
\altaffiliation{NSF Astronomy and Astrophysics Postdoctoral Fellow}
\affiliation{Department of Astronomy \& Astrophysics, University of Chicago, 5640 S Ellis Avenue, Chicago, IL 60637, USA}
\affiliation{Department of Astronomy, University of California, Berkeley, Berkeley, CA 94720, USA}
\author[0000-0002-5259-3974]{Nicholas~Storm}
\affiliation{Max Planck Institute for Astronomy, K\"{o}nigstuhl 17, D-69117 Heidelberg, Germany}
\author[0000-0002-0572-8012]{Vedant~Chandra}
\affiliation{Center for Astrophysics $\mid$ Harvard \& Smithsonian, 60 Garden St, Cambridge, MA 02138, USA}
\author[0000-0001-5761-6779]{Kevin C.\ Schlaufman}
\affiliation{Johns Hopkins University William H.\ Miller III Department of Physics \& Astronomy, 3400 N Charles St, Baltimore, MD 21218, USA}
\affiliation{Tuve Fellow, Carnegie Institution for Science Earth \& Planets Laboratory, 5241 Broad Branch Road NW, Washington, DC 20015, USA}
\author[0000-0002-3481-9052]{Keivan G.\ Stassun}
\affiliation{Department of Physics and Astronomy, Vanderbilt University, Nashville, TN 37235, USA}
\author[0000-0002-3684-1325]{Alexander~Heger}
\affiliation{School of Physics and Astronomy, Monash University, Melbourne, Vic 3800, Australia}
\affiliation{Joint Institute for Nuclear Astrophysics -- Center for Evolution of the Elements (JINA), East Lansing, MI 48824, USA}
\affiliation{Center of Excellence for Astrophysics in Three Dimensions (ASTRO-3D), Stromlo, ACT 2611, Australia}
\author[0000-0002-9048-6010]{Marco~Pignatari}
\affiliation{Konkoly Observatory, Research Centre for Astronomy and Earth Sciences, HUN-REN, Konkoly Thege Mikl\'{o}s \'{u}t 15-17, H-1121 Budapest, Hungary}
\affiliation{CSFK, MTA Centre of Excellence, Budapest, Konkoly Thege Miklós út 15-17, H-1121, Hungary}
\affiliation{E. A. Milne Centre for Astrophysics, University of Hull,     Hull HU6 7RX, UK}
\author[0000-0003-0872-7098]{Adrian~M.~Price-Whelan}
\affiliation{Center for Computational Astrophysics, Flatiron Institute, 162 Fifth Ave, New York, NY 10010, USA}
\author[0000-0002-9908-5571]{Maria~Bergemann}
\affiliation{Max Planck Institute for Astronomy, K\"{o}nigstuhl 17, D-69117 Heidelberg, Germany}
\author[0000-0003-1479-3059]{Guy~S.~Stringfellow}
\affiliation{Center for Astrophysics and Space Astronomy, University of Colorado, 389 UCB, Boulder, CO 80309-0389, USA}
\author[0000-0003-0191-2477]{Carla~Fr\"ohlich}
\affiliation{Department of Physics, North Carolina State University, Raleigh NC 27695, USA}
\affiliation{Joint Institute for Nuclear Astrophysics -- Center for Evolution of the Elements (JINA), East Lansing, MI 48824, USA}
\author[0000-0001-6533-6179]{Henrique~Reggiani}
\altaffiliation{Carnegie Fellow}
\affiliation{The Observatories of the Carnegie Institution for Science, 813 Santa Barbara Street, Pasadena, CA 91101, USA}
\author[0000-0002-5463-6800]{Erika~M.~Holmbeck}
\altaffiliation{NHFP Hubble Fellow}
\affiliation{The Observatories of the Carnegie Institution for Science, 813 Santa Barbara Street, Pasadena, CA 91101, USA}
\author[0000-0002-4818-7885]{Jamie~Tayar}
\affiliation{Department of Astronomy, University of Florida, Bryant Space Science Center, Stadium Road, Gainesville, FL 32611, USA}
\author[0000-0002-3367-2394]{Shivani~P.~Shah}
\affiliation{Department of Astronomy, University of Florida, Bryant Space Science Center, Stadium Road, Gainesville, FL 32611, USA}
\author[0000-0001-9345-9977]{Emily~J.~Griffith}
\altaffiliation{NSF Astronomy and Astrophysics Postdoctoral Fellow}
\affiliation{Center for Astrophysics and Space Astronomy, University of Colorado, 389 UCB, Boulder, CO 80309-0389, USA}
\author[0000-0003-3922-7336]{Chervin~F.~P.~Laporte}
\affiliation{Institut de Ci\`encies del Cosmos (ICCUB), Universitat de Barcelona, Mart\'i i Franqu\`es 1, E-08028 Barcelona, Spain}
\affiliation{Institut d’Estudis Espacials de Catalunya (IEEC), E-08034 Barcelona, Spain}
\author[0000-0003-0174-0564]{Andrew~R.~Casey}
\affiliation{School of Physics \& Astronomy, Monash University, Wellington Road, Clayton 3800, Victoria, Australia}
\affiliation{ARC Centre of Excellence for All Sky Astrophysics in 3 Dimensions(ASTRO 3D), Canberra, ACT 2611, Australia}
\author[0000-0002-1423-2174]{Keith~Hawkins}
\affiliation{Department of Astronomy, The University of Texas at Austin, 2515 Speedway Boulevard, Austin, TX 78712, USA}
\author[0000-0003-1856-2151]{Danny~Horta}
\affiliation{Center for Computational Astrophysics, Flatiron Institute, 162 Fifth Ave, New York, NY 10010, USA}
\author[0000-0003-1697-7062]{William~Cerny}
\affiliation{Department of Astronomy, Yale University, New Haven, CT 06520, USA}
\author[0000-0002-3867-3927]{Pierre~Thibodeaux}
\affiliation{Department of Astronomy \& Astrophysics, University of Chicago, 5640 S Ellis Avenue, Chicago, IL 60637, USA}
\author[0000-0003-0918-7185]{Sam~A.~Usman}
\affiliation{Department of Astronomy \& Astrophysics, University of Chicago, 5640 S Ellis Avenue, Chicago, IL 60637, USA}
\author[0000-0002-7662-5475]{Jo\~ao~A.~S.~Amarante}
\affiliation{Institut de Ciencies del Cosmos (ICCUB), Universitat de Barcelona (IEEC-UB), Martí i Franquès 1, E08028 Barcelona, Spain}
\affiliation{Jeremiah Horrocks Institute, University of Central Lancashire, Preston, PR1 2HE, UK}
\author[0000-0002-1691-8217]{Rachael~L.~Beaton}
\affiliation{Space Telescope Science Institute, 3700 San Martin Drive, Baltimore, MD 21218, USA}
\author[0000-0002-1617-8917]{Phillip~A.~Cargile}
\affiliation{Center for Astrophysics $\mid$ Harvard \& Smithsonian, 60 Garden St, Cambridge, MA 02138, USA}
\author[0000-0003-1269-7282]{Cristina~Chiappini}
\affiliation{Leibniz-Institut für Astrophysik Potsdam (AIP), An der Sternwarte 16, D-14482 Potsdam, Germany}
\author[0000-0002-1590-8551]{Charlie~Conroy}
\affiliation{Center for Astrophysics $\mid$ Harvard \& Smithsonian, 60 Garden St, Cambridge, MA 02138, USA}
\author[0000-0001-7258-1834]{Jennifer~A.~Johnson}
\affiliation{Department of Astronomy, The Ohio State University, 140 W. 18th Avenue, Columbus, OH 43210, USA}
\affiliation{Center for Cosmology and AstroParticle Physics, Ohio State University, 191 West Woodruff Avenue, Columbus, OH 43210, USA}
\author[0000-0001-9852-1610]{Juna~A.~Kollmeier}
\affiliation{The Observatories of the Carnegie Institution for Science, 813 Santa Barbara Street, Pasadena, CA 91101, USA}
\affiliation{Canadian Institute for Theoretical Astrophysics, University of Toronto, Toronto, ON, M5S-98H, Canada}
\author[0000-0002-0389-9264]{Haining~Li}
\affiliation{Key Lab of Optical Astronomy, National Astronomical Observatories, Chinese Academy of Sciences, A20 Datun Road, Chaoyang, Beijing 100012, People’s Republic of China}
\author[0000-0003-3217-5967]{Sarah~Loebman}
\affiliation{Department of Physics, University of California, Merced, 5200 N Lake Road, Merced, CA 95343, USA}
\author[0000-0001-6181-1323]{Georges~Meynet}
\affiliation{Department of Astronomy of Geneva University, Switzerland}
\author[0000-0002-3601-133X]{Dmitry~Bizyaev}
\affiliation{Apache Point Observatory and New Mexico State University, P.O. Box 59, Sunspot, NM, 88349-0059, USA}
\affiliation{Sternberg Astronomical Institute, Moscow State University, Moscow}
\author[0000-0002-8725-1069]{Joel~R.~Brownstein}
\affiliation{Department of Physics and Astronomy, University of Utah, 115 S. 1400 E., Salt Lake City, UT 84112, USA}
\author[0000-0002-3956-2102]{Pramod~Gupta}
\affiliation{Department of Astronomy, University of Washington, Box 351580, Seattle, WA 98195, USA}
\author[0000-0002-6770-2627]{Sean~Morrison}
\affiliation{Department of Astronomy, University of Illinois at Urbana-Champaign, Urbana, IL 61801, USA}
\author[0000-0002-2835-2556]{Kaike~Pan}
\affiliation{Apache Point Observatory and New Mexico State University, P.O. Box 59, Sunspot, NM, 88349-0059, USA}
\author{Solange~V.~Ramirez}
\affiliation{The Observatories of the Carnegie Institution for Science, 813 Santa Barbara Street, Pasadena, CA 91101, USA}
\author[0000-0003-4996-9069]{Hans-Walter~Rix}
\affiliation{Max Planck Institute for Astronomy, K\"{o}nigstuhl 17, D-69117 Heidelberg, Germany}
\author[0000-0003-2486-3858]{Jos\'e~S\'anchez-Gallego}
\affiliation{Department of Astronomy, University of Washington, Box 351580, Seattle, WA 98195, USA}

%% file: abundtable.tex
\begin{deluxetable}{ccrrrrr}
\tablecolumns{7}
\tabletypesize{\footnotesize}
\tablecaption{\label{tab:abunds}Chemical Abundances}
\tablehead{Species & $N$ & $\log \epsilon$ & [X/H] & [X/Fe] & $\sigma$ & $\Delta_{\rm NLTE}$}
\startdata
Li I   &   1 & $ 1.15$ & $+0.10$ & $+1.86$ &  0.12 & \nodata \\
C-H    &   2 & $ 6.07$ & $-2.49$ & $-0.73$ &  0.22 & $+0.10$ \\
N-H    &   1 & $ 5.36$ & $-2.62$ & $-0.86$ & limit & $-0.10$ \\
O I    &   3 & $ 7.23$ & $-1.54$ & $+0.22$ &  0.17 & $-0.06$ \\
Na I   &   2 & $ 2.80$ & $-3.49$ & $-1.73$ &  0.09 & $-0.10$ \\
Mg I   &   8 & $ 5.52$ & $-2.03$ & $-0.27$ &  0.11 & $-0.02$ \\
Al I   &   1 & $ 4.15$ & $-2.28$ & $-0.52$ &  0.50 & $+0.65$ \\
Si I   &   8 & $ 5.80$ & $-1.79$ & $-0.03$ &  0.13 & $-0.17$ \\
K I    &   2 & $ 2.05$ & $-3.09$ & $-1.33$ &  0.14 & $-0.20$ \\
Ca I   &  30 & $ 4.73$ & $-1.64$ & $+0.12$ &  0.13 & $+0.02$ \\
Sc II  &   3 & $-0.06$ & $-3.13$ & $-1.37$ &  0.09 & \nodata \\
Ti II  &  30 & $ 2.56$ & $-2.38$ & $-0.62$ &  0.10 & $+0.12$ \\
V II   &   7 & $ 1.48$ & $-2.41$ & $-0.65$ &  0.10 & \nodata \\
Cr II  &   5 & $ 3.75$ & $-1.99$ & $-0.23$ &  0.12 & \nodata \\
Mn I   &  10 & $ 4.06$ & $-1.46$ & $+0.30$ &  0.06 & $+0.34$ \\
Fe I   & 181 & $ 5.74$ & $-1.76$ & $+0.00$ &  0.13 & $+0.11$ \\
Co I   &   4 & $ 3.37$ & $-1.58$ & $+0.18$ &  0.10 & $+0.11$ \\
Ni I   &  26 & $ 5.04$ & $-1.20$ & $+0.56$ &  0.15 & $+0.31$ \\
Cu I   &   1 & $ 2.64$ & $-1.55$ & $+0.21$ &  0.19 & $+0.35$ \\
Zn I   &   2 & $ 3.17$ & $-1.38$ & $+0.38$ &  0.08 & \nodata \\
Rb I   &   1 & $ 2.61$ & $+0.09$ & $+1.85$ & limit & \nodata \\
Sr II  &   3 & $ 1.85$ & $-1.02$ & $+0.74$ &  0.07 & $+0.03$ \\
Y II   &  24 & $ 1.27$ & $-0.94$ & $+0.82$ &  0.11 & $-0.00$ \\
Zr II  &  15 & $ 1.87$ & $-0.71$ & $+1.05$ &  0.08 & \nodata \\
Nb II  &   1 & $ 1.34$ & $-0.12$ & $+1.64$ & limit & \nodata \\
Mo I   &   1 & $ 0.55$ & $-1.33$ & $+0.43$ &  0.16 & \nodata \\
Ru I   &   1 & $ 0.68$ & $-1.07$ & $+0.69$ &  0.14 & \nodata \\
Rh I   &   1 & $ 0.34$ & $-0.57$ & $+1.19$ & limit & \nodata \\
Pd I   &   1 & $ 0.46$ & $-1.11$ & $+0.65$ &  0.14 & \nodata \\
Ag I   &   1 & $-0.15$ & $-1.09$ & $+0.67$ & limit & \nodata \\
Ba II  &   1 & $-2.33$ & $-4.51$ & $-2.75$ &  0.13 & $+0.18$ \\
Eu II  &   1 & $-1.79$ & $-2.31$ & $-0.55$ & limit & \nodata \\
\enddata
\tablecomments{The \citet{EMagg2022} solar normalization is used. NLTE corrections have already been applied and are shown for reference. For C and N, $\Delta_{\rm NLTE}$ is an estimate for the evolutionary correction. The correction listed is for each element's $\log \epsilon$, so [X/Fe] has an additional correction for [Fe/H] already applied. The abundance uncertainty for all elements is for the relative value [X/Fe], except for [Fe/H] which is the absolute metallicity uncertainty. Upper limits are denoted ``limit'' in the $\sigma$ column.}
\vspace{-3em}
\end{deluxetable}

%% file: nucref.tex
\begin{deluxetable}{lccccll}
\tablecolumns{7}
\tabletypesize{\footnotesize}
\tablecaption{\label{tab:nucref}Nucleosynthesis Yield Grids}
\tablehead{Key & Explosion & Mass Range (\msun) & Energy (B) & Metallicity (Solar) & Comments/Other Parameters}
\startdata
HW10 CC & Piston (S4) & 10-100 & 0.3-2.4 & 0 & Mixing: none to 0.251 \\
HW10 HN & Piston (S4) & 10-100 & 3, 5, 10 & 0 & Mixing: none to 0.251  \\
G18 CC & Piston (S4) & 10-80 & 0.1-2.0 & 0 & Mixing: none to 0.251 \\
G18 HN & Piston (S4) & 10-80 & 2.5-200 & 0 & Mixing: none for $E > 5$B  \\
HW02 & PISN & 140-260 & 9-87 & 0 & Initial mass from He core mass  \\
 &&&&& Energy calculated self-consistently  \\
NKT13 CC & Thermal Bomb & 11-140 & 1 & $0$ & Mixing and fallback  \\
NKT13 HN & Thermal Bomb & 20-140 & 10-71 & $0$ & Mixing and fallback  \\
NKT13 PI & PISN & 140-300 & 16-50 & 0 & Energy calculated self-consistently   \\
LC12 & Kinetic Bomb & 13-80 & N/A & 0 & --  \\
LC18 & Kinetic Bomb & 13-120 & N/A & $10^{-3,-2}$ & Rotation: 0, 150, 300 km/s \\
 &&&&& Includes all forced explosion models  \\
E20 & Engine-driven & 11-31 & 0.3-1.7 & $0, 10^{-4}$ & Energy calculated self-consistently  \\
R18 & Shock and Cool & 12-25 & N/A & $10^{-2.3}$ & Delayed explosions  \\
\enddata
\tablecomments{
Restricted to metallicities $\mbox{[Z/H]} < -1.5$. Other models not searched are NKT13 $Z/Z_\odot = 10^{-1.3,-0.7,-0.4,0,+0.4}$, LC18 $Z/Z_\odot = 0.1,1$, R18 $Z/Z_\odot = 10^{,-1.3,-0.5,-0.2,0}$.
}
\vspace{-5mm}
\end{deluxetable}